\documentclass[aps,prc,twocolumn,superscriptaddress,showpacs,floatfix]{revtex4-1}

\usepackage{graphicx,bm,amsfonts,dcolumn}

\begin{document}

\title{\boldmath Muon spin rotation and relaxation in Pr$_{1-x}$Nd$_x$Os$_4$Sb$_{12}$:\\ Magnetic and superconducting ground states}

\author{D.~E. MacLaughlin}
\affiliation{Department of Physics and Astronomy, University of California, Riverside, California 92521, USA}
\author{P.-C. Ho}
\affiliation{Department of Physics, California State University, Fresno, California 93740, USA}
\author{Lei Shu}
\affiliation{State Key Laboratory of Surface Physics, Department of Physics, Fudan University, Shanghai 200433, People's Republic of China}
\author{O.~O. Bernal}
\affiliation{Department of Physics and Astronomy, California State University, Los Angeles, California 90032, USA}
\author{Songrui Zhao}
\altaffiliation{Current address: Department of Electrical and Computer Engineering, McGill University, Montreal, Quebec, Canada H3A 0E9.}
\affiliation{Department of Physics and Astronomy, University of California, Riverside, California 92521, USA}
\author{A.~A. Dooraghi}
\author{T. Yanagisawa}
\altaffiliation{Current address: Faculty of Science, Hokkaido University, Sapporo, Hokkaido 060-1810, Japan.}
\author{M.~B. Maple}
\affiliation{Department of Physics, University of California, San Diego, La Jolla, California 92093, USA}
\author{R.~H. Fukuda}
\affiliation{Department of Physics, California State University, Fresno, California 93740, USA}

\date{\today}

\begin{abstract}
Muon spin rotation and relaxation ($\mu$SR) experiments have been carried out to characterize magnetic and superconducting ground states in the Pr$_{1-x}$Nd$_x$Os$_4$Sb$_{12}$ alloy series. In the ferromagnetic end compound~NdOs$_4$Sb$_{12}$ the spontaneous local field at positive-muon ($\mu^+$) sites below the ordering temperature~$T_C$ is greater than expected from dipolar coupling to ferromagnetically aligned Nd$^{3+}$ moments, indicating an additional indirect RKKY-like transferred hyperfine mechanism. For $0.45 \le x \le 0.75$, $\mu^+$ spin relaxation rates in zero and weak longitudinal applied fields indicate that static fields at $\mu^+$ sites below $T_C$ are reduced and strongly disordered. We argue this is unlikely to be due to reduction of Nd$^{3+}$ moments, and speculate that the Nd$^{3+}$-$\mu^+$ interaction is suppressed and disordered by Pr doping. In an $x = 0.25$ sample, which is superconducting below $T_c = 1.3$~K, there is no sign of ``spin freezing'' (static Nd$^{3+}$ magnetism), ordered or disordered, down to 25~mK\@. Dynamic $\mu^+$ spin relaxation is strong, indicating significant Nd-moment fluctuations. The $\mu^+$ diamagnetic frequency shift and spin relaxation in the superconducting vortex-lattice phase decrease slowly below $T_c$, suggesting pair breaking and/or possible modification of Fermi-liquid renormalization by Nd spin fluctuations. For $0.25 \le x \le 0.75$, the $\mu$SR data provide evidence against phase separation; superconductivity and Nd$^{3+}$ magnetism coexist on the atomic scale.
\end{abstract}

\maketitle

\section{\label{sec:intro} INTRODUCTION}

Rare-earth-based materials are in many ways ideal for studies of the interaction between superconductivity and magnetism in metals. An example is the filled skutterudite family of isostructural lanthanide intermetallics~\cite{MBHH09}, where the unconventional heavy-fermion superconductor~PrOs$_4$Sb$_{12}$ and its alloys have been the subject of considerable interest~\cite{BFHZ02,MFHY06}. The isomorph~NdOs$_4$Sb$_{12}$ is a ferromagnet with Curie temperature~$T_C \simeq 0.8$~K~\cite{SSNS03,HYBF05,MHYH07}. In order to elucidate the interplay between the superconductivity of PrOs$_4$Sb$_{12}$ and the ferromagnetism of NdOs$_4$Sb$_{12}$, the alloy system~Pr$_{1-x}$Nd$_x$Os$_4$Sb$_{12}$ has been investigated. 

\begin{figure}[ht] 
\includegraphics[clip=,width=8.6cm]{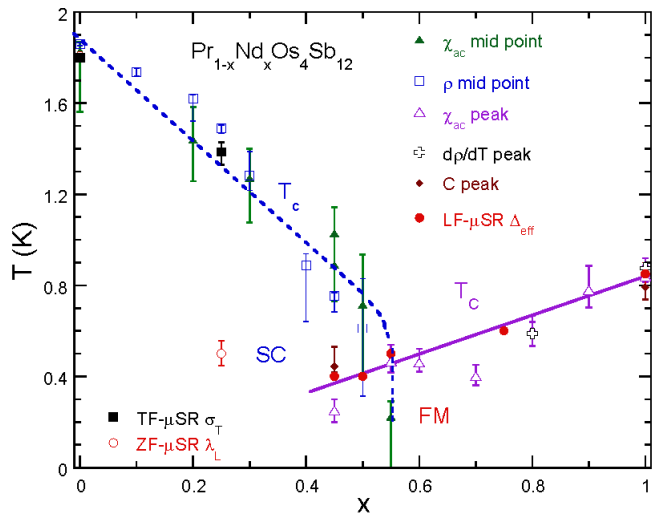}%
\caption{\label{fig:phasediag}(Color online) Phase diagram of Pr$_{1-x}$Nd$_{x}$Os$_4$Sb$_{12}$ from resistivity ($\rho$), ac susceptibility ($\chi_{\text{ac}}$), and specific heat ($C$) measurements~\protect\cite{HYYD11}, LF-$\mu$SR quasistatic relaxation rates $\Delta_{\text{eff}}$ ($0.45 \leq x \leq 1$), and TF-$\mu$SR quasistatic rate $\sigma_T$ and ZF-$\mu$SR dynamic rate~$\lambda_L$ ($x = 0.25$). See also Sec.~\protect\ref{sec:concl}}
\end{figure}

Figure~\ref{fig:phasediag} gives the phase diagram obtained from thermodynamic and transport measurements~\cite{HYYD11}, together with transition temperatures from our muon spin rotation and relaxation ($\mu$SR) data as discussed in Sec.~\ref{sec:res}. 
Superconductivity persists up to $x \simeq 0.5$, and Nd$^{3+}$ ``spin freezing'' (static magnetism, with or without long-range order) appears above $x \simeq 0.45$. This is evidence for competition between superconductivity and magnetism for the ground state of the system, as well as the possibility of a ferromagnetic quantum critical point near $x_{\text{cr}} = 0.4$--0.5. The rate of decrease of the superconducting transition temperature~$T_c$ with Nd concentration is nearly the same in the (Pr$_{1-x}$Nd$_x$)Os$_4$Sb$_{12}$ and Pr(Os$_{1-y}$Ru$_y$)$_4$Sb$_{12}$ alloy systems~\cite{HYBY08}, contrary to the behavior of ``conventional'' superconductors where magnetic impurities are much more effective than nonmagnetic ones in suppressing $T_c$. There is evidence that Nd substitution does not affect the Pr$^{3+}$ CEF splitting nearly as strongly as Ru doping~\cite{MHYH07}. 

A number of aspects of superconductivity and magnetism on the atomic scale are readily accessible to the $\mu$SR technique. Early $\mu$SR studies of PrOs$_4$Sb$_{12}$ revealed two important phenomena: the absence of nodes in the superconducting energy gap~\cite{MSHB02}, and the onset of a spontaneous internal magnetic field below the superconducting transition~temperature~$T_c$~\cite{ATKS03} that indicated broken time-reversal symmetry in the superconducting state. Later $\mu$SR experiments found no change of the muon Knight shift below $T_c$~\cite{HSKO07}, suggesting $p$-wave pairing, resolved a discrepancy between inductive and $\mu$SR penetration-depth measurements~\cite{SMBH09}, ruled out a second phase transition in the superconducting state~\cite{MHMS10c}, and studied the effects of La and Ru doping~\cite{AHSO05,ATSK07,Shu07d,SHAH11}.

This article reports $\mu$SR experiments in the Pr$_{1-x}$Nd$_x$Os$_4$Sb$_{12}$ alloy series, undertaken with the goal of providing microscopic characterization of the ground states of these systems. Their paramagnetic states have also been investigated via $\mu$SR Knight shift measurements, the results of which will be reported in a companion article~\footnote{P.-C. Ho \textit{et al.} (unpublished).}. Three major regions of Nd concentration~$x$ have been found in experiments to date~\cite{HYYD11}: $0.55 \lesssim x \le 1$ (local-moment ferromagnetism, disordered in the alloys), $0.32 \lesssim x \lesssim 0.55$ (coexistence of superconductivity and ferromagnetism?), and $x \lesssim 0.32$ (coexistence of superconductivity with Nd$^{3+}$ local moments). After a brief description of the $\mu$SR experiments in Sec.~\ref{sec:exp}, Sec.~\ref{sec:res} reports and discusses the results from alloys in each of these regions. Conclusions are given in Sec.~\ref{sec:concl}. The principal results of this work are (1)~the discovery of reduced and strongly disordered static magnetic fields below $T_C$ at $\mu^+$ stopping sites in the alloys, (2)~the absence of any static Nd magnetism, ordered or disordered, for $x = 0.25$ above 25 mK, and (3)~microscopic-scale coexistence of superconductivity and magnetism for $0.25 \leq x \leq 0.55$.

\section{\label{sec:exp}EXPERIMENT}

$\mu$SR experiments were carried out at the M15 beam line of the TRIUMF accelerator facility, Vancouver, Canada, using a dilution refrigerator to obtain temperatures in the range~25~mK--3~K\@. Millimeter-sized Pr$_{1-x}$Nd$_x$Os$_4$Sb$_{12}$ single crystals, $x = 0.25$, 0.45, 0.50, 0.55, 0.75, and 1.00, were grown using the self-flux technique~\cite{BSFS01}. Solid solubility is good across the alloy series~\cite{HYBY08, HYYD11}. The samples were characterized by x-ray diffraction and magnetic susceptibility measurements. Each $\mu$SR sample consisted of a mosaic of crystals glued to a $11{\times}16{\times}0.25$~mm 6N Ag plate using GE 7031 varnish, which was attached to the cold finger of the cryostat with a thin layer of Apiezon grease. The crystals were partially oriented, as their (100) faces tended to be parallel to the mounting plate, but the cubic symmetry renders most averages over $\mu^+$ sites (e.g., dipolar field averages) independent of orientation. To ensure isothermal conditions an Ag foil was wrapped around the sample and firmly attached to the cold finger. Magnetic fields in the range~0--200~Oe were applied.

In time-differential $\mu$SR experiments spin-polarized positive muons ($\mu^+$) are implanted in a sample, precess in the local fields at the interstitial $\mu^+$ sites, and decay with a mean lifetime~$\tau_\mu = 2.197~\mu$s via the reaction~$\mu^+ \to e^+ + \nu_e + \overline{\nu}_\mu$. The decay positrons are emitted preferentially in the direction of the $\mu^+$ spin, and are detected using scintillation counters. This asymmetry yields the dependence of the $\mu^+$ spin polarization~$G(t)$ on time~$t$ between implantation and muon decay. Typically ${\sim}10^7$ events are obtained for a single measurement of $G(t)$. 

In transverse-field $\mu$SR (TF-$\mu$SR) the applied field~$\mathbf{H}_T$, which is usually the dominant field at $\mu^+$ sites, is perpendicular to the initial $\mu^+$ spin orientation. The $\mu^+$ spins precess, and the positron count rate for a given direction oscillates with time at a frequency near $\gamma_\mu H_T$~\footnote{Frequency shifts from this value (e.g., the Knight shift in metals) are due to internal fields in the sample.}; here $\gamma_\mu = 2\pi \times 13.553$~kHz/G is the $\mu^+$ gyromagnetic ratio. In a disordered (inhomogeneous) static field the precession frequencies are distributed and the oscillations are damped, or do not appear at all if the disorder is sufficiently strong. 

In longitudinal-field $\mu$SR (LF-$\mu$SR) the $\mu^+$ spins are initially oriented parallel to the applied field~$\mathbf{H}_L$, and hence precess only in local fields generated by the sample such that the resultant field is not parallel to $\mathbf{H}_L$. Zero-field $\mu$SR (ZF-$\mu$SR) can be considered a limiting case of LF-$\mu$SR\@. Here again, any distribution of local-field magnitudes results in damping of the oscillation expected from a unique field, although $G(t)$ functions for LF- and TF-$\mu$SR are quite different. 

In both cases, thermal fluctuations of the $\mu^+$ local fields give rise to dynamic relaxation (the spin-lattice relaxation of NMR)\@. This is the only relaxation mechanism for the $\mu^+$ components parallel to local static fields observed in ZF- and LF-$\mu$SR.

For a given positron counter, the count rate~$N(t)$ is related to $G(t)$ by
\begin{equation} \label{eq:countrate} 
N(t) = N_0\,e^{-t/\tau_\mu}\left[1 + A_0G(t)\right]
\end{equation}
(ignoring uncorrelated background counts), where $N_0$ is the initial count rate, $\tau_\mu$ is the $\mu^+$ lifetime, $A_0$ is the initial count-rate asymmetry (spectrometer-dependent but typically $\sim$0.2), and the polarization function~$G(t)$ is the projection of the time-dependent ensemble $\mu^+$ spin polarization (normalized to 1 at $t = 0$) on the direction to the counter. Various experimental configurations (applied field direction, counter orientations, etc.)\ yield specific forms for $G(t)$, from which information on static and dynamic magnetic properties of the sample can be extracted. Such configurations, and the information they reveal in the case of Pr$_{1-x}$Nd$_x$Os$_4$Sb$_{12}$ alloys, are discussed below in Sec.~\ref{sec:res}\@. More details of the $\mu$SR technique can be found in a number of monographs and review articles~\cite{Sche85,Brew03,Blun99,LKC99,YaDdR11}.

In a ZF-$\mu$SR experiment in a multidomain or polycrystalline magnet with a sufficient number of randomly oriented domains or crystallites, the static local fields~$\mathbf{B}_\mu$ are oriented randomly with respect to the initial direction of the $\mu^+$ spin polarization~$\mathbf{S}_\mu$. At each $\mu^+$ site the $\mu^+$ spin has a nonprecessing ``longitudinal'' component parallel to $\mathbf{B}_\mu$ of magnitude~$S_\mu\cos\theta$, where $\theta$ is the angle between $\mathbf{B}_\mu$ and $\mathbf{S}_\mu$. The projection of this longitudinal component back along the initial $\mu^+$ spin direction (the axis of the spectrometer counter system) is therefore $S_\mu\cos^2\theta$. For randomly oriented $\mathbf{B}_\mu$ the angular average~$\overline{\cos^2\theta} = 1/3$, leading to a ``1/3'' longitudinal contribution to the $\mu^+$ spin polarization in zero applied field~\cite{KuTo67, *HUIN79}. The ``2/3'' transverse contribution oscillates if $B_\mu$ is reasonably constant. Any distribution of field magnitudes (i.e., of $\mu^+$ precession frequencies) leads to damping of the oscillation. This leaves only the 1/3 component at late times, which is time independent if there are no other relaxation mechanisms; otherwise it relaxes dynamically. Hence dynamic and static relaxation can be separated in ZF-$\mu$SR (and weak-LF-$\mu$SR) experiments if the former is sufficiently slow compared to the latter.

In nonzero longitudinal field~$\mathbf{H}_L$ the resultant $\mu^+$ field~$\mathbf{B}_\mu + \mathbf{H}_L$ is no longer randomly oriented. Then $\overline{\cos^2\theta}$ is greater than 1/3, and approaches 1 asymptotically as $\mathbf{H}_L \gg \mathbf{B}_\mu$; the $\mu^+$ spins are ``decoupled''~\cite{KuTo67, *HUIN79} from their local fields. In the absence of dynamic relaxation, one expects a polarization function of the form
\begin{eqnarray} \label{eq:decoup} 
G(t) & = & \left[1 - f_L(H_L)\right]g(H_L, t) + f_L(H_L) , \nonumber \\
1/3 & \le & f_L(H_L) \le 1 ,
\end{eqnarray}
where $g(H_L, t)$ describes the distribution of precession frequencies, and $f_L(H_L)$, the fraction of longitudinal component, is a monotonically increasing function of $H_L$~\cite{Prat07}. 

Since we will be dealing with disordered spin systems in the following sections, we consider $\mu^+$ polarization functions appropriate to such cases. In zero applied field the static Gaussian Kubo-Toyabe ( KT) function~\cite{KuTo67, *HUIN79}
\begin{equation} \label{eq: KT} 
G_{\text{ KT}}(t) = \frac{1}{3} + \frac{2}{3}\left(1 - \Delta^2 t^2\right)\exp\left(-{\textstyle \frac{1}{2}}\Delta^2 t^2\right) 
\end{equation}
describes the $\mu^+$ polarization when the distribution of each Cartesian component of $\mathbf{B}_\mu$ is Gaussian with zero mean and rms width~$\Delta$. The Gaussian  KT function models $\mu^+$ relaxation in dipolar fields from a densely populated lattice of randomly oriented moments, electronic or nuclear, when the moment magnitudes are fixed but their orientations are random on the atomic scale. The Gaussian distributions result from the central limit theorem of statistics, since a given muon is coupled to many lattice moments. For randomly located $\mu^+$ sites in a lattice with a low concentration of moments, however, the wings of the distribution are dominated by single nearby moments, and the distribution of field components becomes nearly Lorentzian~\cite{WaWa74,UYHS85}. This leads to the ``Lorentzian  KT'' ZF-$\mu$SR polarization function~$1/3 + (2/3)(1 - at)\exp(-at)$, where $a$ is the half-width of the field distribution~\cite{Kubo81,UYHS85}. Clearly the form of the polarization function changes appreciably between the dilute and concentrated limits.

Noakes and Kalvius~\cite{NoKa97} have approached the problem of $\mu^+$ relaxation in the intermediate-dilution regime by generalizing the  KT result phenomenologically, assuming that in a moderately diluted lattice the  KT distribution width~$\Delta$ is itself distributed. The physical meaning of this approach is discussed further in Sec.~\ref{sec:concl}. A Gaussian distribution of $\Delta$ with mean~$\Delta_0$ and rms width~$w$ results in the ``Gaussian-broadened Gaussian'' (GbG)  KT ZF-$\mu$SR polarization function~\cite{NoKa97}
\begin{eqnarray} \label{eq:GbG} 
G_{\text{GbG}}(t) & = & \frac{1}{3} + \frac{2}{3} \left( \frac{1 + R^2}{1 + R^2(1 + \Delta_{\text{eff}}^2t^2)} \right)^{3/2} \nonumber \\
& & \times \left(1 - \frac{\Delta_{\text{eff}}^2t^2}{1 + R^2(1 + \Delta_{\text{eff}}^2t^2)} \right) \nonumber \\
& & \times \exp\left(-\frac{\Delta_{\text{eff}}^2t^2}{1 + R^2(1 + \Delta_{\text{eff}}^2t^2)} \right) ,
\end{eqnarray}
where $\Delta_{\text{eff}}^2 = \Delta_0^2 + w^2$ is the effective spin relaxation rate and $R = w/\Delta_0$ is the ratio of the distribution widths. 

\begin{figure}[ht] 
\includegraphics[clip=,width=8.6cm]{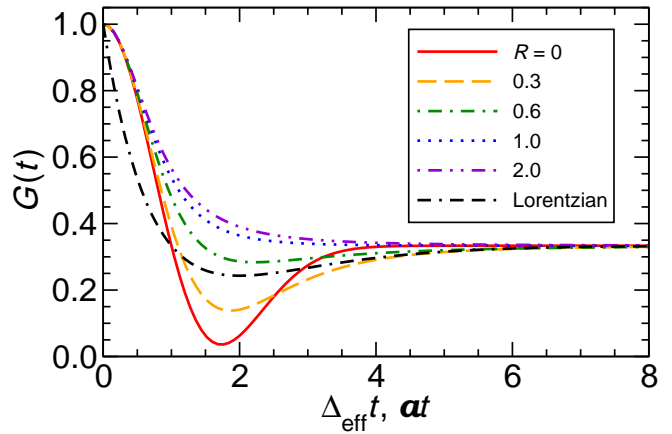}%
\caption{\label{fig: KTfuncs}(Color online) Static zero-field Gaussian-broadened Gaussian and Lorentzian  KT polarization functions (relaxation rates~$\Delta_{\text{eff}}$ and $a$, respectively).}
\end{figure}

GbG  KT polarization functions are shown in Fig.~\ref{fig: KTfuncs} together with the Lorentzian  KT function. 
The ratio~$R$ parameterizes the departure from single-Gaussian  KT behavior~\cite{NoKa97}. For $R = 0$ $G_{\text{GbG}}(t) = G_{\text{ KT}}(t)$, and with increasing $R$ the minimum near $\Delta_0 t = 1$ decreases in depth. For $R \gtrsim 1$ $G_{\text{GbG}}(t)$ does not approach the Lorentzian function, so that $G_{\text{GbG}}(t)$ is not an interpolation function between the Gaussian and Lorentzian limits. Instead, $G_{\text{GbG}}(t)$ becomes monotonic and nearly independent of $R$; for this reason the condition~$R \le 1$ was imposed in fits of Eq.~(\ref{eq:GbG}) to the data. For small $\Delta_{\text{eff}}t$ $G_{\text{GbG}}(t) \simeq 1 - \Delta_{\text{eff}}^2t^2$, independent of $R$ for all $R$.

Often the experimental $\mu$SR asymmetry signal has a contribution from muons that miss the sample and stop in the mounting plate or cryostat cold finger. This plate is usually silver, for good thermal contact and also because Ag nuclear moments are small and $\mu^+$ spin relaxation in Ag is negligible. In the following the plots of ZF- and LF-$\mu$SR data give the polarization function~$G(t)$ for the sample rather than the total asymmetry~$A(t)$. The relation between these quantities is 
\begin{equation} \label{eq:asy2pol} 
G(t) = \frac{A(t)/A_0 - f_{\text{Ag}}}{1 - f_{\text{Ag}}} ,
\end{equation} 
where $A_0$ is the total initial asymmetry and $f_{\text{Ag}}$ is the fraction of muons that stop in the silver.

\section{\label{sec:res}RESULTS AND DISCUSSION} 

\subsection{\label{sec:resNOS} \boldmath NdOs$_4$Sb$_{12}$}

\begin{figure}[ht] 
\includegraphics[clip=,width=8.6cm]{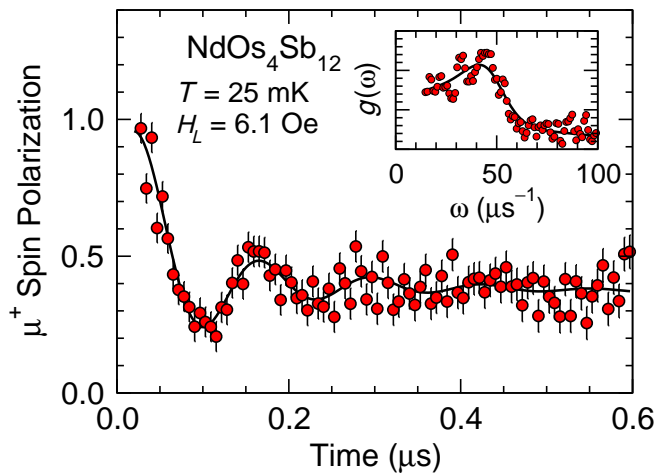}%
\caption{\label{fig:NOS25mKpol}(Color online) Early-time $\mu^+$ spin polarization in NdOs$_4$Sb$_{12}$, $H_L = 6.1$~Oe, $T = 25$~mK\@. Curve: fits of Eq.~(\protect\ref{eq:intpbe}) to the data. Inset: Fourier transforms~$g(\omega)$ of data and fit.}
\end{figure}

In the end compound~NdOs$_4$Sb$_{12}$ the onset of ferromagnetism below the Curie temperature~$T_C \simeq 0.8$~K gives rise to a spontaneous internal field~$\mathbf{B}_\mu$ at $\mu^+$ sites, and therefore to oscillations in ZF-$\mu$SR and weak-LF-$\mu$SR\@. Figure~\ref{fig:NOS25mKpol} shows early-time data taken in a longitudinal field~$H_L = 6.1$~Oe, 
which decouples the $\mu^+$ moments from nuclear dipolar fields above $T_C$ (cf.\ Sec.~\ref{sec:exp}), but has little effect in the ferromagnetic state; the observed internal fields are two orders of magnitude greater than $H_L$ except near $T_C$. The data are well fit by the damped Bessel function
\begin{eqnarray}\label{eq:intpbe} 
G_B^{\text{dmpd}}(t) & = & e^{-\lambda_Lt}\left[(1 - f_L)e^{-\lambda_Tt}J_0(\omega_\mu t+\varphi)+ f_L\right], \nonumber \\
\omega_\mu & = & \gamma_\mu\overline{B}_\mu , 
\end{eqnarray}
where $J_0(x)$ is the zeroth-order cylindrical Bessel function, $\lambda_L$ is the (dynamic) longitudinal spin relaxation rate, $f_L$ is the fraction of longitudinal signal component, $\lambda_T$ describes the damping of the oscillation, $\varphi$ is the initial phase, and $\overline{B}_\mu$ is the dominant value of the $\mu^+$ local-field magnitude distribution. A damped cosine function does not provide as good a fit.

The choice of the Bessel function is motivated by the Fourier-transform relation between the polarization function and the distribution of precession frequencies. In an incommensurate single-$\mathbf{q}$ sinusoidal spin-density wave, the precession frequency distribution is $g(\omega) = 2/\left(\pi\sqrt{\omega_\mu^2 - \omega^2}\right)\quad (0 < \omega < \omega_\mu)$, the Fourier transform of which is the Bessel function~\cite{AFGS95}. The distribution~$g(\omega)$ is characterized by a singularity at $\omega_\mu$ and a broad distribution of lower frequencies. Thus $J_0(\omega_\mu t)$ represents the $\mu^+$ polarization when the experimental frequency distribution has this character (inset of Fig.~\ref{fig:NOS25mKpol}). An actual incommensurate spin-density wave in NdOs$_4$Sb$_{12}$ is ruled out by the considerable evidence for a ferromagnetic ground state~\cite{HYBF05}. Broadening of the singularity is accounted for by damping the Bessel function [Eq.~(\ref{eq:intpbe})]. 

\begin{figure}[ht] 
\includegraphics[clip=,width=8.6cm]{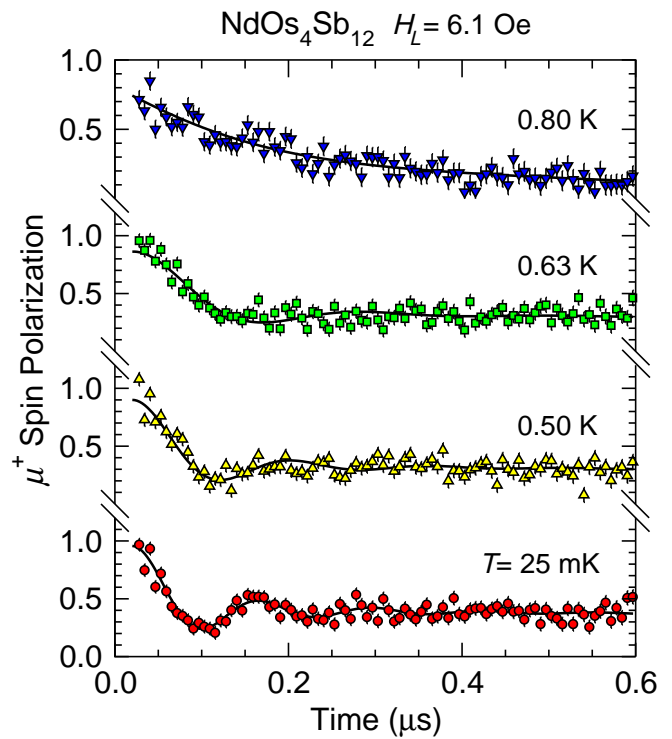}%
\caption{\label{fig:NOS6Gpol}(Color online) Early-time $\mu^+$ spin polarization at various temperatures $\lesssim T_C$ from weak-LF-$\mu$SR in NdOs$_4$Sb$_{12}$, $H_L = 6.1$~Oe\@. Data for $T = 25$~mK from Fig.~\protect\ref{fig:NOS25mKpol}. Curves: fits of Eq.~(\protect\ref{eq:intpbe}) to the data.}
\end{figure}

\begin{figure}[ht] 
\includegraphics[clip=,width=8.6cm]{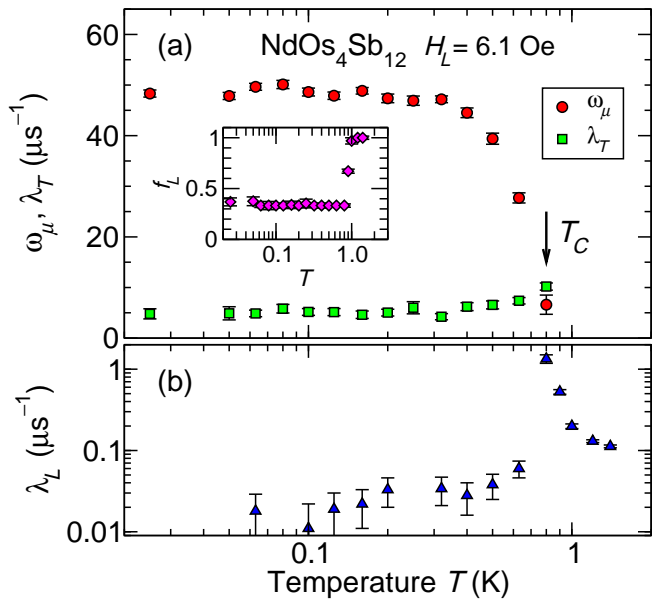}%
\caption{\label{fig:NOSfrqrlx}(Color online) Temperature dependencies of weak-LF-$\mu$SR parameters in NdOs$_4$Sb$_{12}$, $H_L = 6.1$~Oe\@. (a)~Spontaneous $\mu^+$ spin precession frequency~$\omega_\mu$ and static transverse spin relaxation rate~$\lambda_T$. Inset: fraction~$f_L$ of longitudinal component. (b)~Longitudinal spin relaxation rate~$\lambda_L$. Arrow: Curie temperature~$T_C$.}
\end{figure}

$\mu^+$ spin polarizations in NdOs$_4$Sb$_{12}$ at various temperatures below~$T_C$ are shown in Fig.~\ref{fig:NOS6Gpol}, together with damped Bessel-function fits. 
Figure~\ref{fig:NOSfrqrlx} gives the temperature dependencies of $\omega_\mu$, $\lambda_T$, and $\lambda_L$. 
The dominant frequeny~$\omega_\mu(T)$ exhibits an order-parameter-like increase below $T_C$. At 0.80~K $\omega_\mu$ is small but finite (cf.\ Fig.~\ref{fig:NOS6Gpol}), indicating that $T_C$ is slightly higher than this. The damping rate~$\lambda_T$, being considerably larger than $\lambda_L$, is mainly due to static disorder. It is much smaller than $\omega_\mu$ until $T$ approaches $T_C$ from below; here it increases slightly, suggesting a spread of transition temperatures~\cite{[{See, e.g., }]MVBdR94}. The dynamic rate~$\lambda_L(T)$ increases as $T \rightarrow T_C$ from above due to critical slowing down of Nd$^{3+}$ spin fluctuations~\footnote{In the motionally narrowed limit the rate is proportional to the correlation time}, followed by a decrease below $T_C$ as the Nd$^{3+}$ moments freeze. The behavior of all these quantities is that of a conventional ordered magnet. 

The value of $\overline{B}_\mu = \omega_\mu/\gamma_\mu$ at low temperatures is $560 \pm 10$~G\@. Assuming the saturation moment~$M_{\text{sat}} =1.73\mu_B$ found from magnetization isotherms~\cite{HYBF05}, the experimental $\mu^+$-Nd$^{3+}$ coupling constant~$A^{\text{expt}} = \overline{B}_\mu/M_{\text{sat}}$ is $335 \pm 6~\text{G}/\mu_B$. For comparison we have calculated the dipolar coupling tensor~$\mathsf{A}_{\text{dip}}^{\text{calc}}$ from ferromagnetically aligned Nd$^{3+}$ moments~$\mathbf{M}_{\text{Nd}}$, assuming muons stop at the probable $\frac{1}{2},0,0.15$ site as found in PrOs$_4$Sb$_{12}$~\cite{ATKS03}. The field~$\mathbf{B}_{\text{dip}}^{\text{calc}}$ at this site is given by $\mathbf{B}_{\text{dip}}^{\text{calc}} = \mathsf{A}_{\text{dip}}^{\text{calc}}\cdot\mathbf{M}_{\text{Nd}}$. The principal axes of $\mathsf{A}_{\text{dip}}^{\text{calc}}$ are parallel to the crystal axes; the principal-axis values~$A_{\text{dip}}^{\text{calc}}$ and the corresponding values of $B_{\text{dip}}^{\text{calc}}$ for $M_{\text{Nd}} = 1.73\mu_B$ are given in Table~\ref{tab:dipcoup}\@.
\begin{table}[ht] 
\caption{\label{tab:dipcoup} Calculated principal-axis values of the dipolar coupling tensor~$A_{\text{dip}}^{\text{calc}}$ and dipolar fields~$B_{\text{dip}}^{\text{calc}}$ at the probable $\frac{1}{2},0,0.15$ $\mu^+$ site in NdOs$_4$Sb$_{12}$.}
\begin{ruledtabular}
\begin{tabular}{cccc}
Crystal axis & $\mu^+$ site coordinate & $A_{\text{dip}}^{\text{calc}}~(\text{G}/\mu_B)$ & $B_{\text{dip}}^{\text{calc}}~(G)$ 
\\
\hline
$a$ & $\frac{1}{2}$ & $150.6$ & 260.5 \\
$b$ & 0 & $-62.6$ & -108.3 \\
$c$ & 0.15 & $-88.0$ & -152.2 \\
\end{tabular}
\end{ruledtabular}
\end{table}
In general three $\mu^+$ frequencies are expected from the three inequivalent $\mu^+$ sites in the cubic structure for nonero $\mathbf{M}_{\text{Nd}}$. 

All the $B_{\text{dip}}^{\text{calc}}$ are smaller in magnitude than the observed field, so that a significant RKKY-like transferred hyperfine interaction between Nd$^{3+}$ spins and $\mu^+$ spins is necessary to account for the difference. The interaction strength required to do this depends on the orientation of $\mathbf{M}_{\text{Nd}}$, which is not known at present. Broadened unresolved resonances corresponding to the $\sim$370-G spread in principal-axis dipolar fields (${\sim}30~\mu\text{s}^{-1}$ spread in angular precession frequencies) might contribute to the spectral weight below the peak in $g(\omega)$ (inset of Fig.~\ref{fig:NOS25mKpol}). 

\subsection{\label{sec:resPN75OS} \boldmath Pr$_{0.25}$Nd$_{0.75}$NdOs$_4$Sb$_{12}$} 

In Pr$_{1-x}$Nd$_x$Os$_4$Sb$_{12}$ alloys the Pr$^{3+}$ ions are in nonmagnetic crystal-field ground states, resulting in a substitutionally diluted lattice of Nd$^{3+}$ ions~\footnote{But see the discussion in Sec.~\protect\ref{sec:concl}.}. For $x = 0.75$ Pr doping has reduced the magnetic transition temperature~$T_C$ from 0.8~K to $\sim$0.55~K~\cite{HYYD11}. Although there has been no direct confirmation of ferromagnetic order in the diluted alloys, the Nd concentration dependence of the ``Curie'' temperature~$T_C$ and paramagnetic-state properties in the alloys tend smoothly towards their values in NdOs$_4$Sb$_{12}$ as $x(\text{Nd}) \to 1$~\cite{HYYD11}. In Pr$_{0.25}$Nd$_{0.75}$Os$_4$Sb$_{12}$ the paramagnetic Curie-Weiss temperature is positive and $\simeq T_C$.

\begin{figure}[ht] 
\includegraphics[clip=,width=8.6cm]{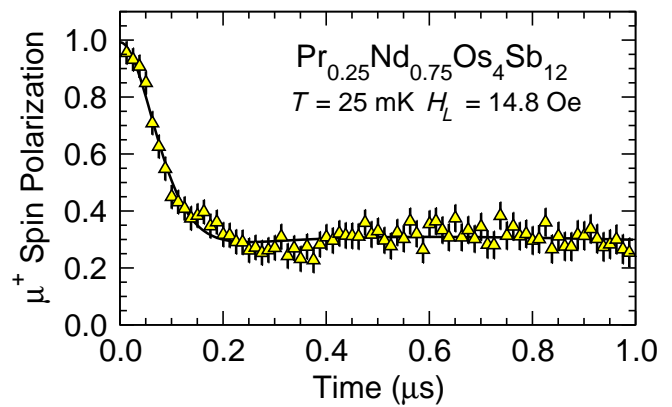}%
\caption{\label{fig:PN75OS25mKpol}(Color online) Early-time $\mu^+$ spin polarization from weak-LF-$\mu$SR in Pr$_{0.25}$Nd$_{0.75}$Os$_4$Sb$_{12}$, $H_L = 14.8$~Oe\@. Curve: fit of Eq.~(\protect\ref{eq:dGbG}) to the data.}
\end{figure}

Figure~\ref{fig:PN75OS25mKpol} shows early-time weak-LF-$\mu$SR spin polarization data from Pr$_{0.25}$Nd$_{0.75}$Os$_4$Sb$_{12}$ at $T = 25$~mK and $H_L = 14.8$~Oe. 
The weak longitudinal field was applied to decouple nuclear dipolar fields above $T_C$ as discussed in Secs.~\ref{sec:exp} and \ref{sec:resNOS}; here $H_L$ is also much smaller than the internal field, at least at low temperatures. 

Compared to data from NdOs$_4$Sb$_{12}$ at 25~mK (Fig.~\ref{fig:NOS6Gpol}) the oscillation is almost completely damped, indicating a broad distribution of local fields. The deep minimum of the Gaussian  KT function (Fig.~\ref{fig: KTfuncs}) is not observed, and we have therefore fit the GbG  KT polarization function~$G_{\text{GbG}}(t)$ [Eq.~(\ref{eq:GbG})] to the data. As in Sec.~\ref{sec:resNOS}, we take dynamic relaxation into account via an exponential damping factor:
\begin{equation} \label{eq:dGbG} 
G_{\text{GbG}}^{\text{dmpd}}(t) = e^{-\lambda_Lt} G_{\text{GbG}}(t) \,.
\end{equation}
The fit of Eq.~(\ref{eq:dGbG}) to the data at 25~mK, the early-time portion of which is shown in Fig.~\ref{fig:PN75OS25mKpol}, is tolerable but not perfect; it is, however, considerably better (reduced $\chi^2 = 1.16$) than that of a number of other candidate functions as follows. 
\begin{enumerate}

\item[(1)] The damped cosine and damped Bessel functions discussed in Sec.~\ref{sec:resNOS} ($\chi^2 = 1.33$ and 1.39, respectively).

\item[(2)] The ``$\delta$-function/Gaussian'' function~\cite{LFGK00}\newline $1/3 + (2/3)[\cos\omega t - (\Delta^2t/\omega)\sin\omega t]\exp(-\frac{1}{2}\Delta^2 t^2)$ ($\chi^2 \simeq 3)$.

\item[(3)] The ``power  KT'' function~\cite{CrCy97}\newline $1/3 + (2/3) \left[1 - (\lambda t)^\beta\right]\exp\left[-(\lambda t)^\beta/\beta\right]$ ($\chi^2 = 1.31$).

\item[(4)] The ``Voigtian  KT'' function~\cite{[{See, e.g., }] MSKG10} \newline $1/3 + (2/3)\left(1 - \lambda t - \Delta^2t^2\right)\exp\left(- \lambda t - \frac{1}{2}\Delta^2t^2\right)$ ($\chi^2 = 1.37$).

\end{enumerate} 
The functions in (1) and (2) above oscillate with defined nonzero frequencies, whereas the last two are phenomenological interpolations between the Gaussian and Lorentzian  KT functions. The poor fits to the oscillating functions are clear evidence that for $x = 0.75$ the spread in $B_\mu$ is considerably greater than the average.

\begin{figure}[ht] 
\includegraphics[clip=,width=8.6cm]{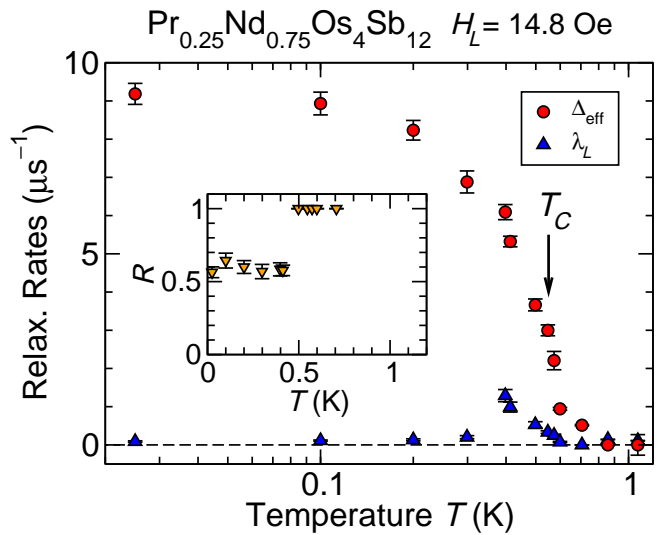}%
\caption{\label{fig:PN75OSrlx}(Color online) Temperature dependencies of parameters from fits of Eq.~(\protect\ref{eq:dGbG}) to weak-LF-$\mu$SR data from Pr$_{0.25}$Nd$_{0.75}$Os$_4$Sb$_{12}$, $H_L = 14.8$~Oe\@. See text for details. Circles: effective $\mu^+$ spin relaxation rate~$\Delta_{\text{eff}}$. Triangles: longitudinal spin relaxation rate~$\lambda_L$. Arrow: magnetic transition temperature~$T_C$ from Ref.~\protect\cite{HYYD11}. Inset: Temperature dependence of the ratio~$R$ of distribution widths.}
\end{figure}

Figure~\ref{fig:PN75OSrlx} gives the temperature dependencies of $\Delta_{\text{eff}}$, $\lambda_L$ and $R$ from fits to Eq.~(\ref{eq:dGbG}). 
Below $T_C$ $\Delta_{\text{eff}}$ increases in an order-parameter-like fashion with decreasing temperature, to $9.2 \pm 0.3~\mu\text{s}^{-1}$ at 25~mK\@. This is $\sim$20\% of $\omega_\mu$ in NdOs$_4$Sb$_{12}$ at the same temperature, to be compared with the much smaller decrease in transition temperature; $T_C(x{=}0.75) \simeq 0.7\,T_C(x{=}1)$. The change in polarization behavior between $x = 0.75$ and 1 is drastic, and the average frequency for $x = 1$ should not be compared in detail to the spread in frequencies for $x = 0.75$. Nevertheless, the difference for this relatively light Pr doping is quite striking. In the neighborhood of $T_C$ $\Delta_{\text{eff}}(T)$ varies rather smoothly without an abrupt transition, suggesting an effect of nonzero $H_L$ and/or an inhomogeneous spread of transition temperatures. The inset of Fig.~\ref{fig:PN75OSrlx} shows that $R$ is nearly constant ($\sim$0.6) at low temperatures, jumping suddenly to $\sim$1 near $T_C$.

The longitudinal rate~$\lambda_L$ behaves similarly in the $x = 1$ and $x = 0.75$ samples, exhibiting a cusp near the magnetic transition. For the $x = 0.75$ sample, however, the cusp occurs at a temperature well below $T_C$, where $\Delta_{\text{eff}}$ has increased to more than 60\% of its low-temperature value. This behavior may also be due to a distribution of transition temperatures. It should be noted, though, that Eq.~(\ref{eq:dGbG}) characterizes the entire polarization function and hence the entire sample volume; the good fits to this function are thus evidence against macroscopic inhomogeneity or phase separation.

\subsection{\label{sec:resPNOS}\boldmath Pr$_{1-x}$Nd$_x$Os$_4$Sb$_{12}$, $x = 0.45$, 0.50, and 0.55} 

\begin{figure}[ht] 
\includegraphics[clip=,width=8.6cm]{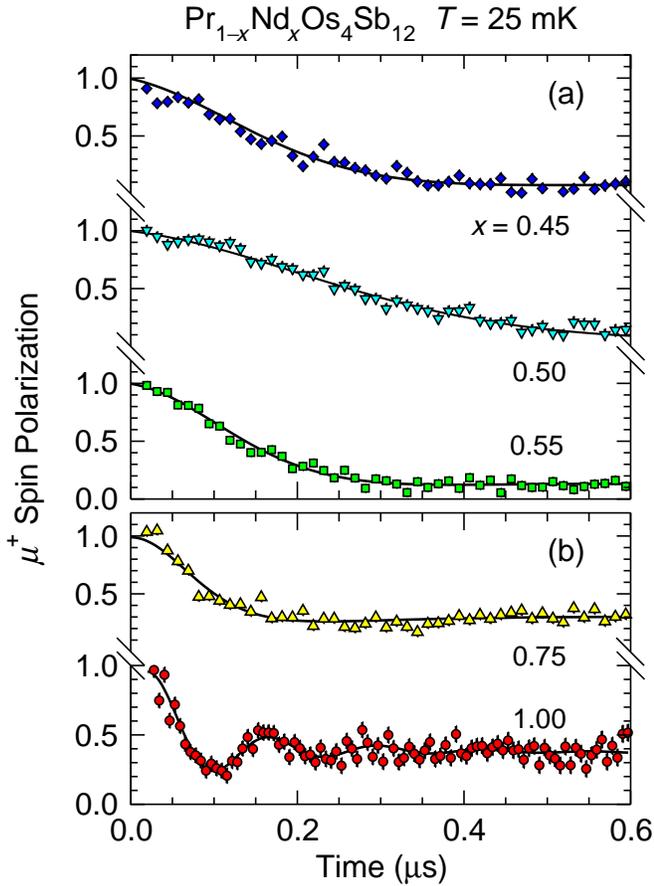}%
\caption{\label{fig:PNOS25mKpol}(Color online) (a)~Early-time $\mu^+$ spin polarization from LF-$\mu$SR in Pr$_{1-x}$Nd$_x$Os$_4$Sb$_{12}$, $x = 0.45$ ($H_L = 15.9$~Oe), 0.50 ($H_L = 16.1$~Oe), and 0.55 ($H_L = 16.3$~Oe), $T = 25$~mK\@. Curves ($x \ne 1$): fits of Eq.~(\protect\ref{eq:dGbG}) to the data. (b)~$x = 0.75$ and 1.00 (data and fits of Figs.~\protect\ref{fig:PN75OS25mKpol} and \protect\ref{fig:NOS6Gpol}, respectively) for comparison.}
\end{figure}

In the concentration range~$0.45 \le x \le 0.55$ the transition from static magnetism to superconductivity is occurring, perhaps with coexistence of the two phases on the microscopic scale and with the possibility of one or more quantum critical points near $x = 0.5$. Figure~\ref{fig:PNOS25mKpol} shows early-time $\mu^+$ spin polarization data for alloys with $x = 0.45$, 0.50, and 0.55. 
The data for $x = 0.75$ and 1.00, discussed above, are repeated for comparison. As is the case for Pr$_{0.25}$Nd$_{0.75}$Os$_4$Sb$_{12}$ (Sec.~\ref{sec:resPN75OS}), the data are fit best by the damped GbG  KT function [Eq.~(\ref{eq:dGbG})]. The two-component structure associated with a distribution of quasistatic local fields is present but in attenuated form, since the damping rate is large in these alloys.  

\begin{figure}[ht] 
\includegraphics[clip=,width=8.6cm]{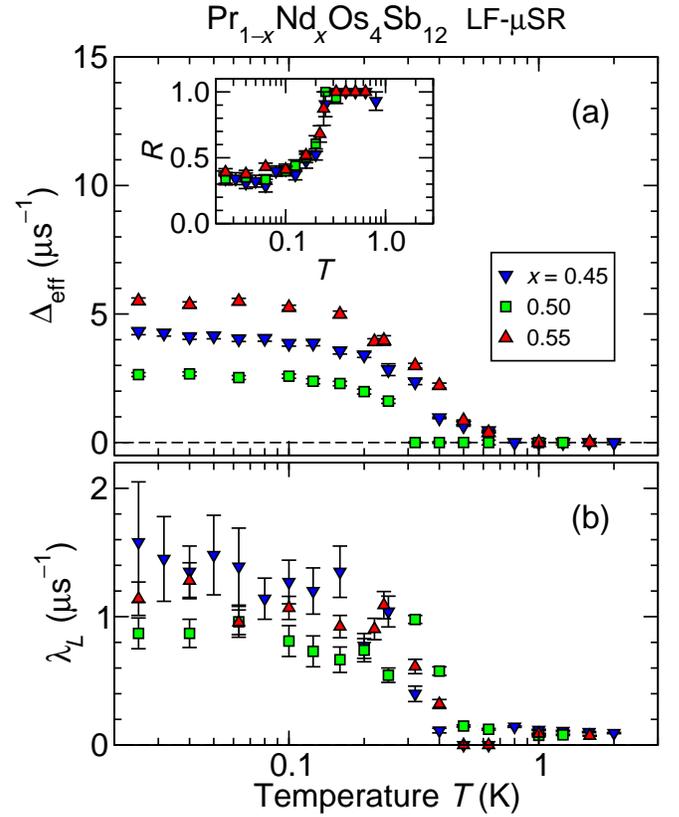}%
\caption{\label{fig:PNOSrlx}(Color online) Temperature dependencies of LF-$\mu$SR parameters in Pr$_{1-x}$Nd$_x$Os$_4$Sb$_{12}$, $x = 0.45$, 0.50, and 0.55, from fits of the damped GbG  KT polarization function [Eq.~(\ref{eq:dGbG})] to the data. See text for details. (a)~Effective $\mu^+$ spin relaxation rates~$\Delta_{\text{eff}}$. Inset: ratios~$R$ of distribution widths. (b)~Longitudinal spin relaxation rates~$\lambda_L$.}
\end{figure}

Figure~\ref{fig:PNOSrlx} gives the temperature dependencies of $\Delta_{\text{eff}}$, $R$, and $\lambda_L$ for $x = 0.45$, 0.50, and 0.55. 
Both rates increase significantly below 0.3--0.5~K\@. The increase of $\Delta_{\text{eff}}$ indicates the onset of a distribution of quasistatic local fields in this temperature range. The transitions occur close to temperatures determined from ac susceptibility measurements~\cite{HYYD11}.

In this Nd concentration range $\Delta_{\text{eff}}$ is suppressed significantly compared to $\omega_\mu$ in NdOs$_4$Sb$_{12}$, continuing the trend found in Pr$_{0.25}$Nd$_{0.75}$Os$_4$Sb$_{12}$ (Sec.~\ref{sec:resPN75OS})\@. It can be seen in Fig.~\ref{fig:PNOSrlx} that $\Delta_{\text{eff}}$ never exceeds ${\sim}5.5~\mu s^{-1}$ at low temperatures. This corresponds to a spread of $\sim$65~Oe in fields at $\mu^+$ sites, about 11\% of the average field in NdOs$_4$Sb$_{12}$. A decrease in $\Delta_{\text{eff}}$ is expected due to the dilution of the Nd$^{3+}$ moment concentration~\cite{Noak91} but not to this extent, as discussed in Sec.~\ref{sec:concl}. The low-temperature values of $\Delta_{\text{eff}}$ do not vary monotonically with $x$, but exhibit a marked minimum for $x = 0.50$. The width ratios~$R(T)$ [inset of Fig.~\ref{fig:PNOSrlx}(a)] behave remarkably similarly for the $x = 0.45$, 0.50, and 0.55 alloys: at low temperatures $R \simeq 0.35$, and then increases toward 1 at $\sim$0.25~K more continuously than for $x = 0.75$ (inset of Fig.~\ref{fig:PN75OSrlx}). The dynamic relaxation rates~$\lambda_L(T)$ [Fig.~\ref{fig:PNOSrlx}(b)] differ considerably from the corresponding data for $x = 0.75$ and 1 in that there is no sign of a peak at or near $T_C$ and the rates remain large down to 25~mK\@. 

\begin{figure}[ht] 
\includegraphics[clip=,width=8.6cm]{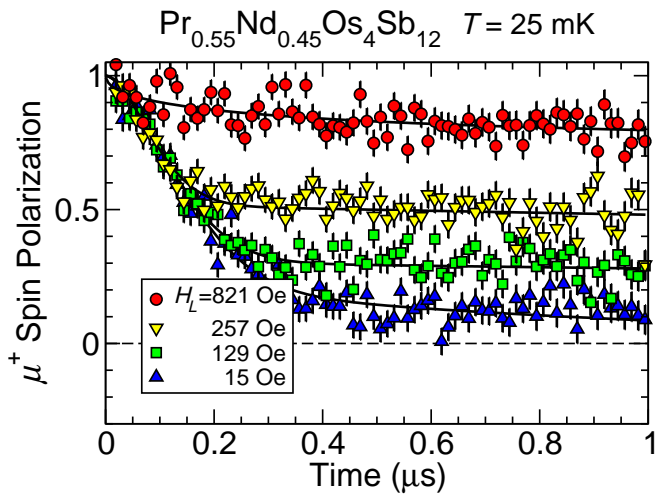}%
\caption{\label{fig:PN45OSpol}(Color online) Early-time LF-$\mu$SR spin polarization functions at $T = 25$~mK in Pr$_{0.55}$Nd$_{0.45}$Os$_4$Sb$_{12}$ at various applied fields.}
\end{figure}

Figure~\ref{fig:PN45OSpol} shows LF-$\mu$SR spin polarization in Pr$_{0.55}$Nd$_{0.45}$Os$_4$Sb$_{12}$ at $T = 25$~mK\@ for $H_L$ in the range 15--821~Oe. 
For intermediate fields, the field independence at early times followed by the increase in the late-time fraction is characteristic of decoupling by the applied field (Sec.~\ref{sec:exp})~\cite{HUIN79}. This is confirmation that the early-time relaxation is static and not dynamic in nature.

\subsection{\label{sec:resPN25OS} \boldmath Pr$_{0.75}$Nd$_{0.25}$Os$_4$Sb$_{12}$}

This alloy is superconducting below a transition temperature~$T_c = 1.3 \pm 0.1$~K from ac susceptibility measurements, compared to ${\sim}1.8$~K for the end compound~PrOs$_4$Sb$_{12}$~\cite{HYBY08,HYYD11}. The suppression of superconductivity by Nd doping has been discussed~\cite{HYYD11} in two alternative scenarios: two-band superconductivity, as found in PrOs$_4$Sb$_{12}$~\cite{MBFS04a,SBMF05}, and the Fulde-Maki multiple pair-breaking theory~\cite{FuMa66}. The main issues addressed by $\mu$SR experiments are therefore the magnetism associated with Nd$^{3+}$ moments, and its effect on the superconductivity of this alloy.

\subsubsection{\label{sec:resPN25OSTF}Transverse Field}

TF-$\mu$SR experiments were carried out in Pr$_{0.75}$Nd$_{0.25}$Os$_4$Sb$_{12}$ in an applied field~$H_T > H_{c1}(T{=}0)$. Damped oscillations were observed both above and below $T_c$. The asymmetry data [$A_0G(t)$ in Eq.~(\ref{eq:countrate})] were fit to a cosine polarization function with combined Gaussian and exponential damping:
\begin{eqnarray} \label{eq:edgtf} 
G_{\text{TF}}(t) & = & \exp\left(-{\textstyle \frac{1}{2}}\sigma_T^2t^2 - \lambda_Tt\right)\cos\left(\omega_\mu t + \varphi\right) , \nonumber \\
\omega_\mu & = & \gamma_\mu\overline{B}_\mu ,
\end{eqnarray}
together with an undamped oscillation from muons that stopped in the silver plate and cold finger. Combined Gaussian and exponential damping is necessary to fit TF-$\mu$SR in PrOs$_4$Sb$_{12}$~\cite{MSHB02}.

\begin{figure}[ht] 
\includegraphics[clip=,width=8.6cm]{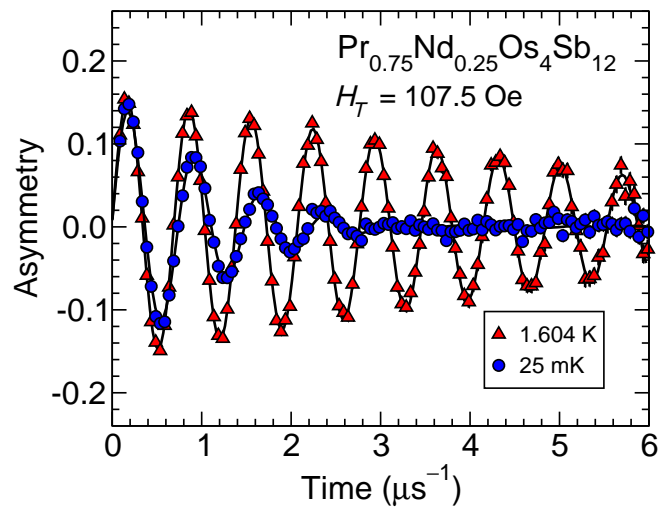}%
\caption{\label{fig:PN25OSwTFasy}(Color online) Weak-TF-$\mu$SR asymmetry data in Pr$_{0.75}$Nd$_{0.25}$Os$_4$Sb$_{12}$, $H_T = 107.5$~Oe, $T = 1.604$~K and 25~mK\@. The signal from muons that stop in the silver cold finger (see text) has been subtracted. Curves: fits of Eq.~(\protect\ref{eq:edgtf}) to the data.} 
\end{figure}

Figure~\ref{fig:PN25OSwTFasy} shows weak-TF-$\mu$SR asymmetry data obtained from Pr$_{0.75}$Nd$_{0.25}$Os$_4$Sb$_{12}$ for $H_T = 107.5$~Oe, at 1.604~K (above $T_c$) and 25~mK (well below $T_c$). 
The signal from muons that did not stop in the sample has been subtracted. It can be seen that at 25~mK the precession frequency decreases and the damping rate increases markedly compared to 1.604~K. 

\begin{figure}[ht] 
\includegraphics[clip=,width=8.6cm]{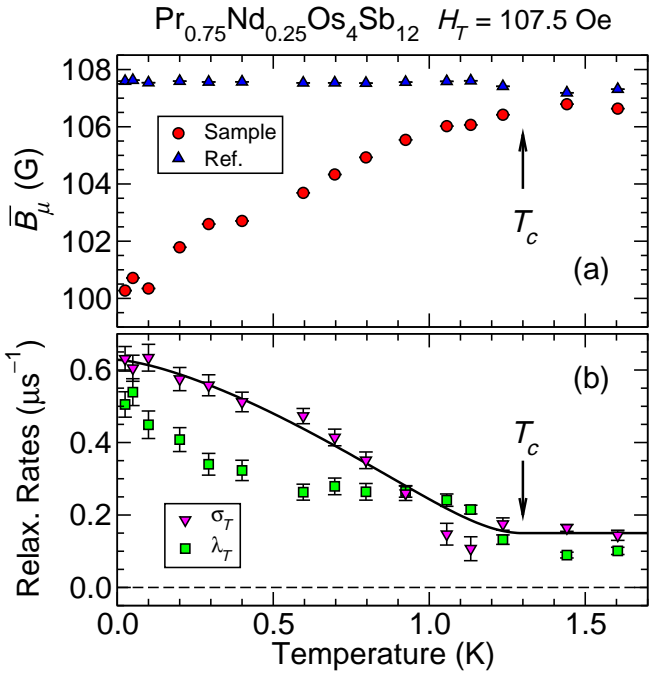}%
\caption{\label{fig:PN25TFparams}(Color online) Temperature dependencies of weak-TF-$\mu$SR parameters in Pr$_{0.75}$Nd$_{0.25}$Os$_4$Sb$_{12}$, $H_T = 107.5$~Oe\@. (a)~Average $\mu^+$ fields~$\overline{B}_\mu$ from $\mu^+$ precession in sample and Ag reference. (b)~Static Gaussian transverse relaxation rate~$\sigma_T$ and exponential transverse relaxation rate~$\lambda_T$. Curve: fit of Eq.~(\protect\ref{eq:sigmaT}) to the data. Arrows: superconducting transition temperature~$T_c$.}
\end{figure}

Figure~\ref{fig:PN25TFparams} gives the temperature dependencies of the parameters obtained by fitting Eq.~(\ref{eq:edgtf}) to the data. 
The value of $\overline{B}_\mu$ for the surrounding silver serves as a reference, and is also plotted in Fig.~\ref{fig:PN25TFparams}(a). The decrease in $\overline{B}_\mu$ and increase in $\sigma_T$ in the superconducting state of the sample are expected from the diamagnetic response and the field distribution in the vortex lattice, respectively. But normally these changes begin much more abruptly just below $T_c$, as is observed in PrOs$_4$Sb$_{12}$~\cite{MSHB02,MHMS10c}.

The curve in Fig.~\ref{fig:PN25TFparams}(b) is a fit of the relation
\begin{equation} \label{eq:sigmaT} 
\sigma_T(T) = \sqrt{\sigma_{sT}^2(T) + \sigma_{nT}^2} , 
\end{equation}
where the temperature dependence of the supercon\-duct\-ing-state Gaussian rate~$\sigma_{sT}$ is modeled by
\begin{equation} \label{eq:sigmas} 
\sigma_{sT}(T) = \left\{ \begin{array}{ll}
\sigma_{sT}(0)\left[ 1 - (T/T_c)^n \right] , & T < T_c \\
0, & T > T_c ,
\end{array} \right.
\end{equation}
and $\sigma_{nT}$ is the normal-state Gaussian rate due to nuclear dipolar fields. The two contributions are added in quadrature because the nuclear dipolar fields that give rise to $\sigma_n$ are randomly oriented and uncorrelated with the vortex-lattice field. The power law used to model $\sigma_{sT}(T)$ [Eq.~(\ref{eq:sigmas})] is merely to indicate the form through the exponent~$n$, and has little physical significance, although $n = 4$ is found in an early two-fluid phenomenology. The fits yield $n = 1.4 \pm 0.2$, much smaller than the usual values (3--4) in conventional superconductors. The exponential rate~$\lambda_T$, which is quite significant, is not approximately constant, as in PrOs$_4$Sb$_{12}$~\cite{SMAT07}, but increases below $T_c$ and exhibits an inflection point at $\sim$0.5~K\@. 

There is no evidence for ``freezing'' of the Nd$^{3+}$ moments. Static magnetism of full Nd$^{3+}$ moments would affect both the muon precession frequency and the damping much more strongly than observed. Further evidence for the lack of spin freezing is discussed below.

\subsubsection{\label{sec:resPN25OSZF}Zero Field}

In most conventional superconductors the only ZF-$\mu$SR relaxation mechanism is provided by nuclear dipolar fields at $\mu^+$ sites. These are not affected by superconductivity, so that the relaxation rate is constant through the transition. The $\mu^+$ spin relaxation is then well fit by the  KT function [Eq.~(\ref{eq: KT})]. In PrOs$_4$Sb$_{12}$, however, it was necessary~\cite{SMAT07} to use the exponentially damped Gaussian  KT function 
\begin{equation} \label{eq:edzfgkt} 
G_{\text{ KT}}^{\text{dmpd}}(t) = e^{-\lambda_Lt}\,G_{\text{ KT}}(t) \,.
\end{equation}
The exponential damping was attributed to dynamic fluctuations of hyperfine-enhanced $^{141}$Pr nuclear moments~\cite{SMAT07}. Equation~(\ref{eq:edzfgkt}) models the case where the local fields fluctuate around static averages rather than around zero~\footnote{This relaxation is longitudinal, since it characterizes the relaxation of components of $\mu^+$ spins parallel to their time-averaged local fields.}. Alternatively, the ``dynamic  KT function''~\cite{KuTo67, *HUIN79}, appropriate when the $\mu^+$ local fields fluctuate as a whole around zero with a fluctuation rate~$\nu$, might be considered.

\begin{figure}[ht] 
\includegraphics[clip=,width=8.6cm]{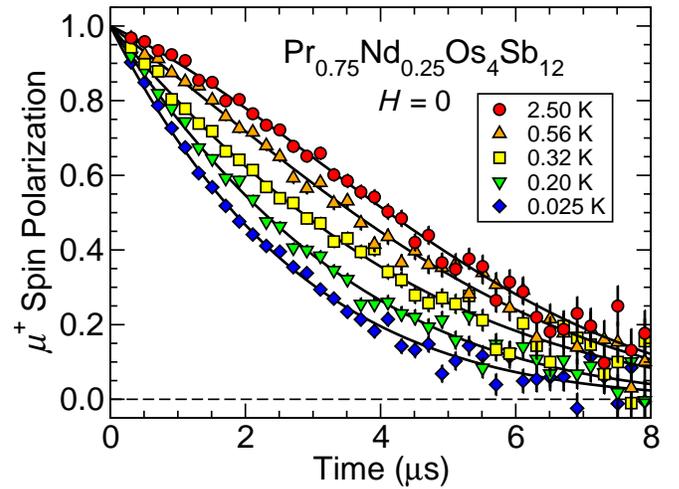}%
\caption{\label{fig:PN25OSZFpol}(Color online) ZF $\mu^+$ spin polarization in Pr$_{0.75}$Nd$_{0.25}$Os$_4$Sb$_{12}$, $25~\text{mK} \le T \le 2.50$~K\@. Curves: fits of Eq.~(\protect\ref{eq:edzfgkt}) to the data.}
\end{figure}

Figure~\ref{fig:PN25OSZFpol} shows the ZF $\mu^+$ spin polarization in Pr$_{0.75}$Nd$_{0.25}$Os$_4$Sb$_{12}$ at a number of temperatures in the range~25~mK--2.50~K\@. 
As in PrOs$_4$Sb$_{12}$, the data can be well fit with Eq.~(\ref{eq:edzfgkt}) at all temperatures, with a significant contribution by the exponential damping. The dynamic  KT function scenario seems unlikely, assuming that $\nu$ decreases with decreasing temperature. If $\nu \ll \Delta$ the overall relaxation of the dynamic  KT function would decrease monotonically with decreasing $\nu$~\cite{KuTo67, *HUIN79}, contrary to observation (Fig.~\ref{fig:PN25OSZFpol}). If $\nu \gg \Delta$ the relaxation is in the motionally narrowed limit and the rate increases with decreasing $\nu$, but then the relaxation would be exponential at all temperatures~\cite{KuTo67, *HUIN79}, again contrary to observation. We conclude that the damped static  KT function is the better choice.

\begin{figure}[ht] 
\includegraphics[clip=,width=8.6cm]{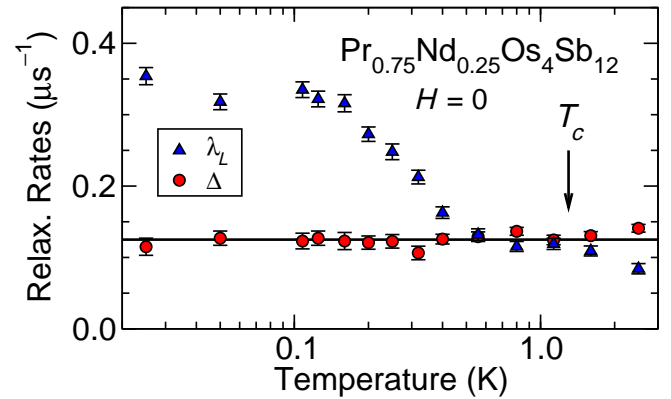}
\caption{\label{fig:PN25ZFrates}(Color online) Temperature dependencies of static Gaussian  KT $\mu^+$ spin relaxation rate~$\Delta$ and dynamic longitudinal spin relaxation rate~$\lambda_L$ from ZF-$\mu$SR in Pr$_{0.75}$Nd$_{0.25}$Os$_4$Sb$_{12}$. Line: average~$\overline{\Delta(T)}$. Arrow: superconducting transition temperature~$T_c$.}
\end{figure}

The temperature dependencies of $\Delta$ and $\lambda_L$ are shown in Fig.~\ref{fig:PN25ZFrates}.
In contrast to the transverse-field Gaussian rate~$\sigma_T$ [Fig.~\ref{fig:PN25TFparams}(b)], the zero-field Gaussian rate~$\Delta$ shows almost no temperature dependence through the superconducting transition down to 25~mK\@. The normal-state values of $\Delta$ and $\sigma_T$ are similar and essentially the same as in PrOs$_4$Sb$_{12}$, consistent with their attribution to Sb nuclear dipolar fields~\cite{ATKS03,SMAT07}. The dynamic rate~$\lambda_L \simeq 0.1~\mu\text{s}^{-1}$ in the normal state is also essentially the same as in PrOs$_4$Sb$_{12}$, but increases dramatically below $\sim$0.4~K, indicating the onset of a new relaxation mechanism at this temperature. 

We note that the observed constant~$\Delta$ here and $\Delta_{\text{eff}}$ in Pr$_{0.75}$Nd$_{0.25}$Os$_4$Sb$_{12}$ are in contrast to the increase of $\Delta$ observed in PrOs$_4$Sb$_{12}$ below $T_c$ and attributed to broken time-reversal symmetry in the superconducting state~\cite{ATKS03,SHAH11}. Nd doping appears to have restored time-reversal symmetry in the superconductivity of the alloys. This seems somewhat paradoxical, since in the BCS theory spin scattering of conduction electrons breaks time-reversed Cooper pairs.

\begin{figure}[ht] 
\includegraphics[clip=,width=8.6cm]{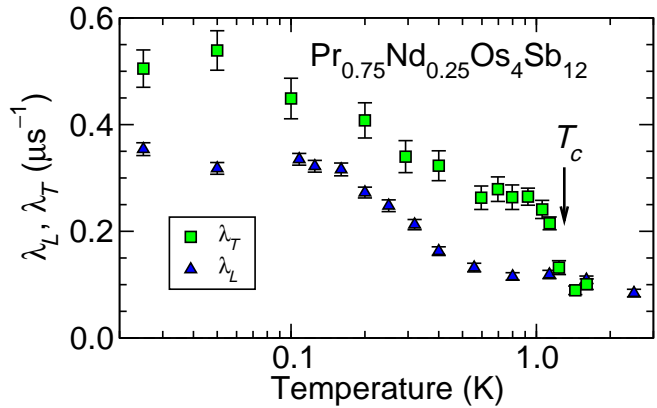}%
\caption{\label{fig:PN25lambda}(Color online) Comparison of spin relaxation rates~$\lambda_L$ ($H = 0$) and $\lambda_T$ ($H_T = 107.5$~Oe) in Pr$_{0.75}$Nd$_{0.25}$Os$_4$Sb$_{12}$. Arrow: superconducting transition temperature~$T_c$.}
\end{figure}

In Fig.~\ref{fig:PN25lambda} the temperature dependencies of $\lambda_L$ for $H = 0$ and $\lambda_T$ for $H_T = 107.5$~Oe are compared. 
Although both relaxation rates show upturns in the region 0.4--0.5~K, they are clearly different: $\lambda_T$ is larger than $\lambda_L$, and exhibits a jump at $T_c$~\footnote{$\lambda_L$ is a purely dynamic rate that describes ``lifetime broadening,'' and thus is always a lower bound on $\lambda_T$.}. These features are discussed in more detail in Sec.~\ref{sec:concl}.

\subsubsection{\label{sec:resPN25OS25mK}$T = 25$~mK}

\begin{figure}[ht] 
\includegraphics[clip=,width=8.6cm]{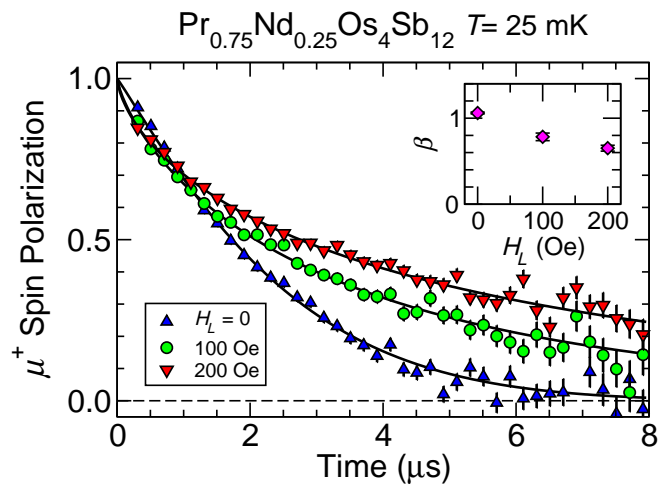}%
\caption{\label{fig:PN25OS25mKpol}(Color online) $\mu^+$ spin polarization in Pr$_{0.75}$Nd$_{0.25}$Os$_4$Sb$_{12}$, $T = 25$~mK\@. Curves: fits of Eq.~(\protect\ref{eq:pedzfgkt}) ($H_L = 0$) and Eq.~(\ref{eq:genexp}) ($H_L = 10$ and 200~Oe) to the data. Inset: power exponent~$\beta$ vs longitudinal field.}
\end{figure}

Further characterization of the unusual $\mu^+$ spin relaxation behavior observed in ZF-$\mu$SR was obtained from LF-$\mu$SR experiments in Pr$_{0.75}$Nd$_{0.25}$Os$_4$Sb$_{12}$ at $H_L = 100$ and 200~Oe, $T = 25$~mK\@. These fields are an order of magnitude larger than that needed to decouple the field distribution width of $\sim$1~Oe determined from the ZF relaxation rates~\cite{KuTo67, *HUIN79}, and hence would completely suppress the relaxation if it were due solely to static fields. But Fig.~\ref{fig:PN25OS25mKpol} shows that $\mu^+$ spin relaxation in these fields is considerable, even though reduced by the field. 

The LF polarization function is subexponential, i.e., exhibits more upward curvature than an exponential function, for $H_L \ge 100$~Oe. This signals an inhomogeneous distribution of dynamic relaxation rates, with the initial slope of $G(t)$ giving the average rate, and slowly relaxing regions dominating at late times after the rapidly relaxing regions have lost their spin polarization. The power exponential function
\begin{equation} \label{eq:genexp} 
G^{\text{pe}}(t) = \exp\left[-(\lambda_L t)^\beta\right]
\end{equation}
fits the LF data well (curves in Fig.~\ref{fig:PN25OS25mKpol} for $H_L = 100$ and 200~Oe), yielding a so-called stretched exponential~\cite{John06} ($\beta < 1$, inset of Fig.~\ref{fig:PN25OS25mKpol}). Although this form is phenomenological and has no theoretical significance, it often fits subexponential data well. For consistency, power-exponential damping of the  KT function
\begin{equation} \label{eq:pedzfgkt} 
G_{\text{ KT}}^{\text{pe dmpd}}(t) = G^{\text{pe}}(t)\,G_{\text{ KT}}(t)
\end{equation}
was used to fit the ZF data. The resulting value of $\beta$ for $H_L = 0$ is close to 1 (inset of Fig.~\ref{fig:PN25OS25mKpol}), justifying our previous use of simple exponential damping in this case, but $\beta$ decreases with increasing field. The value of $\lambda_L$ in the stretched exponential is not the average rate but a rough characterization of the relaxation; $1/\lambda_L$ is the time at which $G^{\text{pe}}(t)$ has decreased to $1/e$ of its initial value. Different values of $\lambda_L$ should not be compared if $\beta$ is also varying, since then the shape of the polarization function is changing.

\section{\label{sec:concl}CONCLUSIONS}

\paragraph{Phase diagram.} Magnetic transition temperatures obtained from weak-LF-$\mu$SR in Pr$_{1-x}$Nd$_x$Os$_4$Sb$_{12}$, $0.45 \leq x \leq 1$ and the superconducting transition from TF-$\mu$SR in the $x = 0.25$ sample are in good agreement with previous results~\cite{HYYD11} (Fig.~\ref{fig:phasediag}). In the latter sample, however, there is no sign of a transition in other data at $\sim$0.5~K, where a marked increase in ZF-$\mu$SR dynamic relaxation is seen (Fig.~\ref{fig:PN25ZFrates}). Specific-heat and other measurements on a $x = 0.25$ sample in this temperature range would be desirable.

For all Nd concentrations, the $\mu$SR spectra (after subtraction of the background Ag signal) exhibit either a single component (TF-$\mu$SR) or the two-component structure that is intrinsic to LF-$\mu$SR\@. Furthermore, the total sample asymmetry is observed not to change within $\sim$5\% at the transitions, magnetic or superconducting; there is no ``lost asymmetry.'' Thus there is no evidence in our data for phase separation anywhere in the phase diagram.

\paragraph{\label{sec:conclNOS} Magnetism in NdOs$_4$Sb$_{12}$.} The weak-LF-$\mu$SR data from NdOs$_4$Sb$_{12}$ reveal a fairly standard ferromagnet, with evidence of disorder from the damped Bessel-function form of the $\mu^+$ polarization function below $T_C$. A distribution of $B_\mu$ magnitudes is necessary for this damping; a random distribution of field directions with fixed field magnitude results in an undamped oscillation in the polarization function. The most likely origin of this distribution is disorder in Nd$^{3+}$ orientations in the ferromagnetic state. We note, however, that as discussed below much broader spreads of $B_\mu$ are observed in Pr-diluted alloys, the mechanism for which might play a role in NdOs$_4$Sb$_{12}$. 

The observation of a peak at $T_C$ in the dynamic relaxation rate indicates critical slowing down of Nd$^{3+}$ spin fluctuations as the transition is approached from above, and thus is evidence that the transition is second order. Critical slowing down is somewhat in disagreement with the conclusion that the transition is mean-field-like~\cite{HYBF05}, however, since in that case the critical region around $T_C$ would be expected to be quite small.

\paragraph{\label{sec:conclPN75OS} Magnetism in Pr$_{0.25}$Nd$_{0.75}$Os$_4$Sb$_{12}$: reduced moments or reduced coupling strengths?} The markedly reduced value of $\Delta_{\text{eff}}$ (and hence $B_\mu$) and the striking difference in polarization function compared to NdOs$_4$Sb$_{12}$ (Sec.~\ref{sec:resPN75OS}), are the salient results of $\mu$SR experiments in Pr$_{0.25}$Nd$_{0.75}$Os$_4$Sb$_{12}$. Noakes~\cite{Noak91,Noak99} has reported Monte Carlo calculations of $B_\mu$ distributions under various conditions of random site dilution, moment direction, and moment magnitude. For fixed moment magnitudes and random moment orientations he found that with increased dilution the $\mu^+$ spin polarization function retains the Gaussian  KT form with its deep minimum (Fig.~\ref{fig: KTfuncs}) down to moment concentration $\sim$0.5. This is clearly not observed in Pr$_{0.25}$Nd$_{0.75}$Os$_4$Sb$_{12}$ (Fig.~\ref{fig:PN75OS25mKpol}). Noakes showed~\cite{Noak99} that an inhomogeneous distribution of moment magnitudes can give rise to the shallower minimum exhibited by the GbG  KT function [Eq.~(\ref{eq:GbG}), $R \gtrsim 0.3$; cf.\ Fig.~\ref{fig: KTfuncs}]~\cite{NoKa97}.

The usual mechanism for local-moment suppression in metals is the Kondo effect, which for $4f$ ions is normally observed only in Ce-, Yb-, and (very occasionally) Pr-based materials. This restriction to the ends of the lanthanide series is well understood~\cite{[{See, e.g., }] Hews93}: the energy difference between the $4f$ level and the Fermi energy increases with increasing atomic number. This decreases the effective $sD\mathrm{-}f$ exchange interaction due to $4f$-conduction electron hybridization at the Fermi surface, to which the Kondo temperature is exponentially sensitive, and thus quenches the Kondo effect~\footnote{A similar argument involving $4f$ holes applies at the high-$Z$ end of the lanthanide series.}. An inhomogeneous Kondo effect may be involved in Pt- and Cu-doped CeNiSn, where the GbG polarization function also fits the ZF-$\mu$SR data well~\cite{KFTK97}.  To the authors' knowledge, however, Kondo screening has never been observed in Nd compounds.

 As noted above, the Pr$^{3+}$ ions are in nonmagnetic crystal-field ground states, but it should not be assumed that they are magnetically inert. Exchange interactions, mediated by a RKKY-like indirect mechanism, might admix magnetic excited CEF states into the nonmagnetic Pr$^{3+}$ ground state so that they contribute to $B_\mu$. It is hard to see how $B_\mu$ would be reduced by this effect, however. 

$\Delta_{\text{eff}}^2$ is a sum of terms, each of which is proportional to the squared product of the  static Nd$^{3+}$ moment magnitude and the Nd$^{3+}$-$\mu^+$ coupling strength~\cite{Abra61}. An alternative mechanism for variation and reduction of $B_\mu$ might invoke a negative contribution of admixed Pr$^{3+}$ states to the indirect Nd$^{3+}$-$\mu^+$ coupling. Variation of moment magnitudes and coupling strengths seem indistinguishable because of the product form, and the latter has the advantage of avoiding a mysterious reduction of Nd$^{3+}$ moments. As far as we know, such a mechanism has not been addressed theoretically. 

\paragraph{\label{sec:conclPN4575OS} Magnetism in Pr$_{1-x}$Nd$_x$Os$_4$Sb$_{12}$, $0.45 \le x \le 0.55$.} The $\mu$SR experiments in this concentration range are dominated by effects of Nd$^{3+}$ magnetism; there is no sign in the data of a superconducting contribution to $B_\mu$. The superconducting state with broken time-reversal symmetry found in PrOs$_4$Sb$_{12}$ is not found in these alloys.

\begin{figure}[ht] 
\includegraphics[clip=,width=8.6cm]{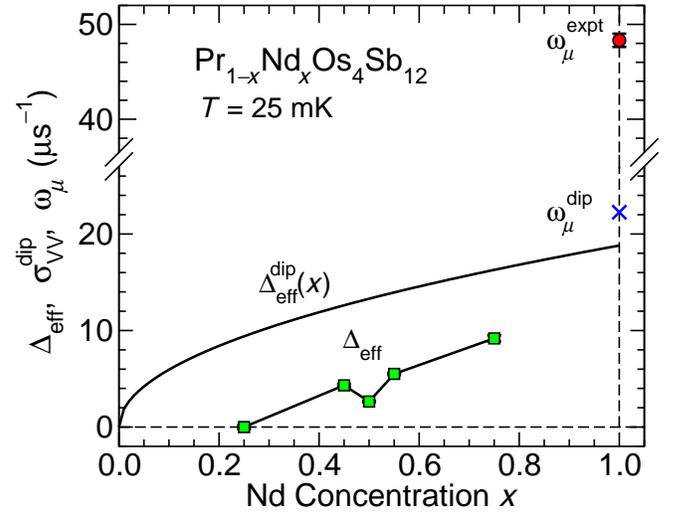}%
\caption{\label{fig:PN25OS25mKDeltaeff}(Color online) ZF- and weak-LF-$\mu$SR GbG effective spin relaxation rates~$\Delta_{\text{eff}}$ ($x < 1$, squares) and experimental $\mu^+$ precession frequency~$\omega_\mu^{\text{expt}}$, $x = 1$ (circle), vs Nd concentration~$x$ in Pr$_{1-x}$Nd$_x$Os$_4$Sb$_{12}$, $T = 25$~mK\@. Cross: calculated $\mu^+$ precession frequency~$\omega_\mu^{\text{dip}}$ in maximum NdOs$_4$Sb$_{12}$ dipolar local field, assuming saturation Nd$^{3+}$ moment~$\mu_{\text{sat}} = 1.73\mu_B$ (Ref.~\protect\cite{HYBF05}) and uniform (ferromagnetic) moment alignment. Curve: calculated $\Delta_{\text{eff}}^{\text{dip}}(x)$ from dipolar $\mu^+$ local fields [= Van Vleck rate~$\sigma_{\text{VV}}^{\text{dip}}(x)$, see text], assuming the saturation Nd$^{3+}$ moment of $1.73\mu_B$ and random moment orientations.}
\end{figure}

The pattern of good fits to the GbG polarization function with reduced values of $\Delta_{\text{eff}}$ is continued. Figure~\ref{fig:PN25OS25mKDeltaeff} shows the dependence of $\Delta_{\text{eff}}$ on Nd concentration for the alloys (including $x = 0.25$, where $\Delta_{\text{eff}}$ vanishes, and $x = 0.75$), 
together with the experimental precession frequency~$\omega_\mu^{\text{expt}}$ and the maximum calculated precession frequency~$\omega_\mu^{\text{dip}}$ assuming dipolar interactions only (Sec.~\ref{sec:resNOS}) in NdOs$_4$Sb$_{12}$. 

In general the initial curvature of $G(t)$ is related to the $\mu^+$ precession frequency distribution: $G_{\text{ZF}}(t) = 1 - \sigma_{\text{VV}}^2t^2 + \cdots$, where $\sigma_{\text{VV}}^2$ is the Van Vleck second moment of the high-field  resonance line~\cite{HUIN79,Abra61}. This result is independent of the functional form of the static field distribution, as long as the second moment is defined~\footnote{The second moment diverges for the Lorentzian distribution.}.  For the GbG  KT function $G_{\text{GbG}}(t) = 1 - \Delta_{\text{eff}}^2t^2 + \cdots$~\cite{NoKa97} (Sec.~\ref{sec:exp}), so that $\Delta_{\text{eff}} = \sigma_{\text{VV}}$. Furthermore, in a diluted lattice of randomly oriented moments the second moment is proportional to the moment concentration~\footnote{Ref.~\protect\cite{Abra61}, Chap.~IV.}: $\sigma_{\text{VV}}(x)/\sigma_{\text{VV}}(x{=}1) = \sqrt{x}$. A lattice-sum calculation of $\sigma_{\text{VV}}$ for dipolar coupling of 1.73$\mu_B$ Nd$^{3+}$ moments to $\mu^+$ spins at $0,\frac{1}{2},0.13$ sites in NdOs$_4$Sb$_{12}$ yields $\sigma_{\text{VV}}^{\text{dip}}(x{=}1) = 18.8~\mu\text{s}^{-1}$. The curve in Fig.~\ref{fig:PN25OS25mKDeltaeff} gives $\Delta_{\text{eff}}^{\text{dip}}(x) = \sigma_{\text{VV}}^{\text{dip}}(x) = 18.8\sqrt{x}$ ($\mu\text{s}^{-1}$), a lower limit for the expected relaxation rate since an additional RKKY-based $\mu^+$-Nd$^{3+}$ transferred hyperfine interaction is present (Sec.~\ref{sec:resNOS}). The observed values of $\Delta_{\text{eff}}$ fall significantly below this curve.

At 25~mK $\Delta_{\text{eff}}$ depends significantly on $x$, exhibiting a marked minimum at $x = 0.50$ (Fig.~\ref{fig:PN25OS25mKDeltaeff}). This behavior is also hard to understand: in general $\Delta_{\text{eff}}$ would be expected to track $T_C$, i.e., $\Delta_{\text{eff}}$ should decrease monotonically with decreasing $x$ in the neighborhood of $x_{\text{cr}}$ (Fig.~\ref{fig:phasediag}). The minimum suggests that the moment suppression is related to an approach to quantum criticality at $x \simeq 0.5$. 

The behavior of the dynamic $\mu^+$ relaxation at low temperatures differs significantly between the $x = 0.45$, 0.50, and 0.55 alloys and the higher-concentration materials. In the former the rate~$\lambda_L$ remains large as $T \rightarrow 0$ [Fig.~\ref{fig:PNOSrlx}(b)] rather than decreasing with decreasing temperature below $T_C$ as in the $x = 0.75$ alloy and NdOs$_4$Sb$_{12}$ (Figs.~\ref{fig:PN75OSrlx} and \ref{fig:NOSfrqrlx}, respectively). This behavior signals the persistence of spin fluctuations to low temperatures. Normally the amplitude of thermal fluctuations decreases at low temperatures with the decreasing population of low-lying excitations such as spin waves. Persistent spin dynamics (PSD) such as seen here are often found in geometrically frustrated spin liquids~\cite{[{See, e.g., }] LMM11p79}, but only occasionally in systems with long-range order~\cite{[{E.g., }]PZZA12}. PSD may be related to the absence of spin freezing in the $x = 0.25$ alloy, discussed below. They might also be considered as a source of the reduced static $\mu^+$ fields (Fig.~\ref{fig:PN25OS25mKDeltaeff}), except that this is also seen in the $x = 0.75$ alloy where PSD are absent (Fig.~\ref{fig:PN75OSrlx}).

\paragraph{\label{sec:conclPN25OS} Pr$_{0.75}$Nd$_{0.25}$Os$_4$Sb$_{12}$: superconductivity and no spin freezing.} As noted above, the observation that the TF-$\mu$SR asymmetry does not change in the superconducting state (Fig.~\ref{fig:PN25OSwTFasy}) is strong evidence that the latter occupies essentially all of the $x = 0.25$ sample; there is no evidence for separation of superconducting and magnetic phases.

From Fig.~\ref{fig:PN25ZFrates} the ZF-$\mu$SR Gaussian  KT rate~$\Delta(T)$ does not deviate from its average $\overline{\Delta(T)} = 0.125 \pm 0.008~\mu\text{s}^{-1}$ by more than ${\sim}0.02~\mu\text{s}^{-1}$, i.e., the $\mu^+$ field does not change by more than $\sim$0.8~G over the entire temperature range. This is even more compelling evidence against static Nd magnetism than the TF-$\mu$SR data (Fig.~\ref{fig:PN25TFparams}), since frozen Nd$^{3+}$ moments~${\sim}1\mu_B$ would produce local fields much greater than 0.8~G\@. An estimate of such a field can be obtained from the calculated rms dipolar relaxation rate (curve in Fig.~\ref{fig:PN25OS25mKDeltaeff}), which is ${\sim}10~\mu\text{s}^{-1} \simeq \gamma_\mu \times 120$~G for $x = 0.25$. Since this calculation uses the measured saturation moment of $1.73\mu_B$/Nd ion in NdOs$_4$Sb$_{12}$, the data suggest an upper limit on any static moment of ${\sim}10^{-2}\mu_B$ in Pr$_{0.75}$Nd$_{0.25}$Os$_4$Sb$_{12}$.

$\mu$SR is sensitive to the onset of static magnetism independent of the degree of order; the technique is as applicable to spin glasses as to ordered magnets.~\cite{Sche85,Brew03,Blun99,LKC99,YaDdR11}. The absence of evidence for static Nd$^{3+}$ magnetism, ordered or disordered, down to 25~mK in Pr$_{0.75}$Nd$_{0.25}$Os$_4$Sb$_{12}$ is perhaps the most surprising result of this study. If $T_C$ were to scale as the Nd concentration the freezing temperature would be $\sim$0.2~K; its suppression by at least an order of magnitude is extraordinary. 

In Pr$_{0.75}$Nd$_{0.25}$Os$_4$Sb$_{12}$ the values at $T = 0$ of the diamagnetic shift ~$\overline{B}_\mu/H_T - 1 \simeq -0.07$ and Gaussian relaxation rate~$\sigma_s = 0.61 \pm 0.03~\mu\text{s}^{-1}$ are of the same order of magnitude as in the end compound~PrOs$_4$Sb$_{12}$ for comparable $H_T$. In the London limit (penetration depth~$\lambda \gg \text{coherence}$ length~$\xi$) the rms width~$\overline{\Delta B^2}$ of the vortex-lattice field distribution in a conventional superconductor is related to the London penetration depth~$\lambda$ by $\overline{\Delta B^2} = 0.00371\Phi_0^2/\lambda^4$, where $\Phi_0$ is the flux quantum~\cite{Bran88}. Assuming that $\gamma_\mu (\overline{\Delta B^2})^{1/2}$ is approximated by the TF-$\mu$SR Gaussian relaxation rate $\sigma_s$, at least near $T = 0$~\footnote{This approximation is well obeyed in PrOs$_4$Sb$_{12}$ (Ref.~\protect\cite{SMBH09}).}, the zero-temperature fit value of $\sigma_s$ yields $\lambda = 4190$~\AA, compared to 3610~\AA\ in PrOs$_4$Sb$_{12}$. 

The increase below $T_c$ of the diamagnetic shift and $\sigma_s$ are much faster in 
superconducting PrOs$_4$Sb$_{12}$~\cite{MHMS10c} than for $x = 0.25$ (Fig.~\ref{fig:PN25TFparams}). One possible mechanism for this behavior is strong pair-breaking by Nd$^{3+}$ spins. A more speculative possibility is that the mass renormalization characteristic of heavy-fermion superconductivity~\cite{Varm85}, which affects the temperature dependence of the penetration depth~\cite{VMS-R86}, is strongly modified by interaction with fluctuating Nd moments.

The increases of both $\lambda_T(T)$ and $\lambda_L(T)$ below $T_c$ (Fig.~\ref{fig:PN25lambda}) indicate a marked effect of superconductivity on the Nd$^{3+}$ spin dynamics. As noted in Sec.~\ref{sec:resNOS}, an increase in relaxation rate with decreasing temperature signals slowing down of spin fluctuations. The observation that $\lambda_T(T) > \lambda_L(T)$ is not surprising, since a contribution to relaxation from a static field distribution is possible in TF-$\mu$SR but not for the 1/3 component of ZF-$\mu$SR\@. The extra transverse relaxation may reflect inhomogeneity in the vortex lattice. The increases of both $\lambda_T(T)$ and $\lambda_L(T)$ below 0.4--0.5~K may be due to a further reduction of the Nd spin fluctuation rate at a crossover or transition, although there is no anomaly in $H_{c2}(T)$ in Pr$_{0.75}$Nd$_{0.25}$Os$_4$Sb$_{12}$ at this temperature~\cite{HYYD11}.

\paragraph{Summary.} $\mu$SR data from the Pr$_{1-x}$Nd$_x$Os$_4$Sb$_{12}$ alloy series reveal a disordered reduction by Pr doping of the spontaneous static $\mu^+$ local field due to static Nd$^{3+}$-ion magnetism that is is well beyond that expected from dilution (Fig.~\ref{fig:PN25OS25mKDeltaeff}). Kondo-like reduction of Nd$^{3+}$ moments is highly unlikely, suggesting by default an effect involving the Nd$^{3+}$-$\mu^+$ coupling strength. The absence of static moments (upper limit ${\sim}10^{-2}\mu_B$) in a $x = 0.25$ sample down to 25~mK may be due to a related suppression of indirect Nd-Nd exchange. The origins of these phenomena remain unclear, and future work is called for to elucidate their mechanisms.
 
\begin{acknowledgments}
We are grateful to the Centre for Material and Molecular Sciences, TRIUMF, for facility support during these experiments. Thanks to J.~M. Mackie and B. Samsonuk for assistance with data taking and analysis, and to Y. Aoki and W. Higemoto for useful discussions. This research was supported by the U.S. National Science Foundation, Grants No.~0422674 and No.~0801407 (UC Riverside), No.~0802478 and No.~1206553 (UC San Diego), No.~1104544 (CSU Fresno), and 1105380 (CSU Los Angeles), by the U.S. Department of Energy, Grant No.~DE-FG02-04ER46105 (UC San Diego), by the National Natural Science Foundation of China (11204041), the Natural Science Foundation of Shanghai, China (12ZR1401200), and the Research Fund for the Doctoral Program of Higher Education of China (2012007112003) (Shanghai), and by the Japanese MEXT (Hokkaido).
\end{acknowledgments}


\begin{thebibliography}{10}%
\makeatletter
\providecommand \@ifxundefined [1]{%
 \ifx #1\undefined \expandafter \@firstoftwo
 \else \expandafter \@secondoftwo
\fi
}%
\providecommand \@ifnum [1]{%
 \ifnum #1\expandafter \@firstoftwo
 \else \expandafter \@secondoftwo
\fi
}%
\providecommand \enquote [1]{``#1''}%
\providecommand \bibnamefont  [1]{#1}%
\providecommand \bibfnamefont [1]{#1}%
\providecommand \citenamefont [1]{#1}%
\providecommand\href[0]{\@sanitize\@href}%
\providecommand\@href[1]{\endgroup\@@startlink{#1}\endgroup\@@href}%
\providecommand\@@href[1]{#1\@@endlink}%
\providecommand \@sanitize [0]{\begingroup\catcode`\&12\catcode`\#12\relax}%
\@ifxundefined \pdfoutput {\@firstoftwo}{%
 \@ifnum{\z@=\pdfoutput}{\@firstoftwo}{\@secondoftwo}%
}{%
 \providecommand\@@startlink[1]{\leavevmode}%
 \providecommand\@@endlink[0]{}%
}{%
 \providecommand\@@startlink[1]{%
  \leavevmode
  \pdfstartlink
   attr{/Border[0 0 1 ]/H/I/C[0 1 1]}%
   user{/Subtype/Link/A<</Type/Action/S/URI/URI(#1)>>}%
  \relax
 }%
 \providecommand\@@endlink[0]{\pdfendlink}%
}%
\providecommand \url  [0]{\begingroup\@sanitize \@url }%
\providecommand \@url [1]{\endgroup\@href {#1}{\urlprefix}}%
\providecommand \urlprefix [0]{URL }%
\providecommand \Eprint[0]{\href }%
\@ifxundefined \urlstyle {%
  \providecommand \doi [1]{doi:\discretionary{}{}{}#1}%
}{%
  \providecommand \doi [0]{doi:\discretionary{}{}{}\begingroup
  \urlstyle{rm}\Url }%
}%
\providecommand \doibase [0]{http://dx.doi.org/}%
\providecommand \Doi[1]{\href{\doibase#1}}%
\providecommand \bibAnnote [3]{%
  \BibitemShut{#1}%
  \begin{quotation}\noindent
    \textsc{Key:}\ #2\\\textsc{Annotation:}\ #3%
  \end{quotation}%
}%
\providecommand \bibAnnoteFile [2]{%
  \IfFileExists{#2}{\bibAnnote {#1} {#2} {\input{#2}}}{}%
}%
\providecommand \typeout [0]{\immediate \write \m@ne }%
\providecommand \selectlanguage [0]{\@gobble}%
\providecommand \bibinfo [0]{\@secondoftwo}%
\providecommand \bibfield [0]{\@secondoftwo}%
\providecommand \translation [1]{[#1]}%
\providecommand \BibitemOpen[0]{}%
\providecommand \bibitemStop [0]{}%
\providecommand \bibitemNoStop [0]{.\EOS\space}%
\providecommand \EOS [0]{\spacefactor3000\relax}%
\providecommand \BibitemShut [1]{\csname bibitem#1\endcsname}%
\bibitem{MBHH09}%
  \BibitemOpen
  \bibfield{author}{%
  \bibinfo {author} {\bibfnamefont{M.~B.}\ \bibnamefont{Maple}}, \bibinfo
  {author} {\bibfnamefont{R.~E.}\ \bibnamefont{Baumbach}}, \bibinfo {author}
  {\bibfnamefont{J.~J.}\ \bibnamefont{Hamlin}}, \bibinfo {author}
  {\bibfnamefont{P.-C.}\ \bibnamefont{Ho}}, \bibinfo {author}
  {\bibfnamefont{L.}~\bibnamefont{Shu}}, \bibinfo {author}
  {\bibfnamefont{D.~E.}\ \bibnamefont{MacLaughlin}}, \bibinfo {author}
  {\bibfnamefont{Z.}~\bibnamefont{Henkie}}, \bibinfo {author}
  {\bibfnamefont{R.}~\bibnamefont{Wawryk}}, \bibinfo {author}
  {\bibfnamefont{T.}~\bibnamefont{Cichorek}},\ and\ \bibinfo {author}
  {\bibfnamefont{A.}~\bibnamefont{Pietraszko}},\ }%
  in\ \emph{\bibinfo {booktitle} {Properties and Applications of Thermoelectric
  Materials}},\ \bibinfo {series and number} {NATO Science for Peace and
  Security Series B: Physics and Biophysics},\ \bibinfo {editor} {edited by\
  \bibinfo {editor} {\bibfnamefont{V.}~\bibnamefont{Zlati\'c}}\ and\ \bibinfo
  {editor} {\bibfnamefont{A.~C.}\ \bibnamefont{Hewson}}}\ (\bibinfo {publisher}
  {Springer Netherlands},\ \bibinfo {year} {2009})\ pp.\ \bibinfo {pages}
  {1--18}%
  \bibAnnoteFile{NoStop}{MBHH09}%
\bibitem{BFHZ02}%
  \BibitemOpen
  \bibfield{author}{%
  \bibinfo {author} {\bibfnamefont{E.~D.}\ \bibnamefont{Bauer}}, \bibinfo
  {author} {\bibfnamefont{N.~A.}\ \bibnamefont{Frederick}}, \bibinfo {author}
  {\bibfnamefont{P.-C.}\ \bibnamefont{Ho}}, \bibinfo {author}
  {\bibfnamefont{V.~S.}\ \bibnamefont{Zapf}},\ and\ \bibinfo {author}
  {\bibfnamefont{M.~B.}\ \bibnamefont{Maple}},\ }%
  \bibfield{journal}{%
  \bibinfo {journal} {Phys. Rev. B}\ }%
  \textbf{\bibinfo {volume} {65}},\ \bibinfo {pages} {100506(R)} (\bibinfo
  {year} {2002})%
  \bibAnnoteFile{NoStop}{BFHZ02}%
\bibitem{MFHY06}%
  \BibitemOpen
  \bibfield{author}{%
  \bibinfo {author} {\bibfnamefont{M.}~\bibnamefont{Maple}}, \bibinfo {author}
  {\bibfnamefont{N.}~\bibnamefont{Frederick}}, \bibinfo {author}
  {\bibfnamefont{P.-C.}\ \bibnamefont{Ho}}, \bibinfo {author}
  {\bibfnamefont{W.}~\bibnamefont{Yuhasz}},\ and\ \bibinfo {author}
  {\bibfnamefont{T.}~\bibnamefont{Yanagisawa}},\ }%
  \bibfield{journal}{%
  \bibinfo {journal} {J. Supercond. Novel Magn.}\ }%
  \textbf{\bibinfo {volume} {19}},\ \bibinfo {pages} {299} (\bibinfo {year}
  {2006})%
  \bibAnnoteFile{NoStop}{MFHY06}%
\bibitem{SSNS03}%
  \BibitemOpen
  \bibfield{author}{%
  \bibinfo {author} {\bibfnamefont{H.}~\bibnamefont{Sato}}, \bibinfo {author}
  {\bibfnamefont{H.}~\bibnamefont{Sugawara}}, \bibinfo {author}
  {\bibfnamefont{T.}~\bibnamefont{Namiki}}, \bibinfo {author}
  {\bibfnamefont{S.~R.}\ \bibnamefont{Saha}}, \bibinfo {author}
  {\bibfnamefont{S.}~\bibnamefont{Osaki}}, \bibinfo {author}
  {\bibfnamefont{T.~D.}\ \bibnamefont{Matsuda}}, \bibinfo {author}
  {\bibfnamefont{Y.}~\bibnamefont{Aoki}}, \bibinfo {author}
  {\bibfnamefont{Y.}~\bibnamefont{Inada}}, \bibinfo {author}
  {\bibfnamefont{H.}~\bibnamefont{Shishido}}, \bibinfo {author}
  {\bibfnamefont{R.}~\bibnamefont{Settai}},\ and\ \bibinfo {author}
  {\bibfnamefont{Y.}~\bibnamefont{Onuki}},\ }%
  \bibfield{journal}{%
  \bibinfo {journal} {J. Phys.:\ Condens. Matter}\ }%
  \textbf{\bibinfo {volume} {15}},\ \bibinfo {pages} {S2063} (\bibinfo {year}
  {2003})%
  \bibAnnoteFile{NoStop}{SSNS03}%
\bibitem{HYBF05}%
  \BibitemOpen
  \bibfield{author}{%
  \bibinfo {author} {\bibfnamefont{P.-C.}\ \bibnamefont{Ho}}, \bibinfo {author}
  {\bibfnamefont{W.~M.}\ \bibnamefont{Yuhasz}}, \bibinfo {author}
  {\bibfnamefont{N.~P.}\ \bibnamefont{Butch}}, \bibinfo {author}
  {\bibfnamefont{N.~A.}\ \bibnamefont{Frederick}}, \bibinfo {author}
  {\bibfnamefont{T.~A.}\ \bibnamefont{Sayles}}, \bibinfo {author}
  {\bibfnamefont{J.~R.}\ \bibnamefont{Jeffries}}, \bibinfo {author}
  {\bibfnamefont{M.~B.}\ \bibnamefont{Maple}}, \bibinfo {author}
  {\bibfnamefont{J.~B.}\ \bibnamefont{Betts}}, \bibinfo {author}
  {\bibfnamefont{A.~H.}\ \bibnamefont{Lacerda}}, \bibinfo {author}
  {\bibfnamefont{P.}~\bibnamefont{Rogl}},\ and\ \bibinfo {author}
  {\bibfnamefont{G.}~\bibnamefont{Giester}},\ }%
  \bibfield{journal}{%
  \bibinfo {journal} {Phys. Rev. B}\ }%
  \textbf{\bibinfo {volume} {72}},\ \bibinfo {pages} {094410} (\bibinfo {year}
  {2005})%
  \bibAnnoteFile{NoStop}{HYBF05}%
\bibitem{MHYH07}%
  \BibitemOpen
  \bibfield{author}{%
  \bibinfo {author} {\bibfnamefont{M.~B.}\ \bibnamefont{Maple}}, \bibinfo
  {author} {\bibfnamefont{Z.}~\bibnamefont{Henkie}}, \bibinfo {author}
  {\bibfnamefont{W.~M.}\ \bibnamefont{Yuhasz}}, \bibinfo {author}
  {\bibfnamefont{P.-C.}\ \bibnamefont{Ho}}, \bibinfo {author}
  {\bibfnamefont{T.}~\bibnamefont{Yanagisawa}}, \bibinfo {author}
  {\bibfnamefont{T.~A.}\ \bibnamefont{Sayles}}, \bibinfo {author}
  {\bibfnamefont{N.~P.}\ \bibnamefont{Butch}}, \bibinfo {author}
  {\bibfnamefont{J.~R.}\ \bibnamefont{Jeffries}},\ and\ \bibinfo {author}
  {\bibfnamefont{A.}~\bibnamefont{Pietraszko}},\ }%
  \bibfield{journal}{%
  \bibinfo {journal} {J. Magn. Magn. Mater.}\ }%
  \textbf{\bibinfo {volume} {310}},\ \bibinfo {pages} {182} (\bibinfo {year}
  {2007})%
  \bibAnnoteFile{NoStop}{MHYH07}%
\bibitem{HYYD11}%
  \BibitemOpen
  \bibfield{author}{%
  \bibinfo {author} {\bibfnamefont{P.-C.}\ \bibnamefont{Ho}}, \bibinfo {author}
  {\bibfnamefont{T.}~\bibnamefont{Yanagisawa}}, \bibinfo {author}
  {\bibfnamefont{W.~M.}\ \bibnamefont{Yuhasz}}, \bibinfo {author}
  {\bibfnamefont{A.~A.}\ \bibnamefont{Dooraghi}}, \bibinfo {author}
  {\bibfnamefont{C.~C.}\ \bibnamefont{Robinson}}, \bibinfo {author}
  {\bibfnamefont{N.~P.}\ \bibnamefont{Butch}}, \bibinfo {author}
  {\bibfnamefont{R.~E.}\ \bibnamefont{Baumbach}},\ and\ \bibinfo {author}
  {\bibfnamefont{M.~B.}\ \bibnamefont{Maple}},\ }%
  \bibfield{journal}{%
  \bibinfo {journal} {Phys. Rev. B}\ }%
  \textbf{\bibinfo {volume} {83}},\ \bibinfo {pages} {024511} (\bibinfo {year}
  {2011})%
  \bibAnnoteFile{NoStop}{HYYD11}%
\bibitem{HYBY08}%
  \BibitemOpen
  \bibfield{author}{%
  \bibinfo {author} {\bibfnamefont{P.-C.}\ \bibnamefont{Ho}}, \bibinfo {author}
  {\bibfnamefont{T.}~\bibnamefont{Yanagisawa}}, \bibinfo {author}
  {\bibfnamefont{N.~P.}\ \bibnamefont{Butch}}, \bibinfo {author}
  {\bibfnamefont{W.~M.}\ \bibnamefont{Yuhasz}}, \bibinfo {author}
  {\bibfnamefont{C.~C.}\ \bibnamefont{Robinson}}, \bibinfo {author}
  {\bibfnamefont{A.~A.}\ \bibnamefont{Dooraghi}},\ and\ \bibinfo {author}
  {\bibfnamefont{M.~B.}\ \bibnamefont{Maple}},\ }%
  \bibfield{journal}{%
  \bibinfo {journal} {Physica B}\ }%
  \textbf{\bibinfo {volume} {403}},\ \bibinfo {pages} {1038} (\bibinfo {year}
  {2008})%
  \bibAnnoteFile{NoStop}{HYBY08}%
\bibitem{MSHB02}%
  \BibitemOpen
  \bibfield{author}{%
  \bibinfo {author} {\bibfnamefont{D.~E.}\ \bibnamefont{MacLaughlin}}, \bibinfo
  {author} {\bibfnamefont{J.~E.}\ \bibnamefont{Sonier}}, \bibinfo {author}
  {\bibfnamefont{R.~H.}\ \bibnamefont{Heffner}}, \bibinfo {author}
  {\bibfnamefont{O.~O.}\ \bibnamefont{Bernal}}, \bibinfo {author}
  {\bibfnamefont{B.-L.}\ \bibnamefont{Young}}, \bibinfo {author}
  {\bibfnamefont{M.~S.}\ \bibnamefont{Rose}}, \bibinfo {author}
  {\bibfnamefont{G.~D.}\ \bibnamefont{Morris}}, \bibinfo {author}
  {\bibfnamefont{E.~D.}\ \bibnamefont{Bauer}}, \bibinfo {author}
  {\bibfnamefont{T.~D.}\ \bibnamefont{Do}},\ and\ \bibinfo {author}
  {\bibfnamefont{M.~B.}\ \bibnamefont{Maple}},\ }%
  \bibfield{journal}{%
  \bibinfo {journal} {Phys. Rev. Lett.}\ }%
  \textbf{\bibinfo {volume} {89}},\ \bibinfo {pages} {157001} (\bibinfo {year}
  {2002})%
  \bibAnnoteFile{NoStop}{MSHB02}%
\bibitem{ATKS03}%
  \BibitemOpen
  \bibfield{author}{%
  \bibinfo {author} {\bibfnamefont{Y.}~\bibnamefont{Aoki}}, \bibinfo {author}
  {\bibfnamefont{A.}~\bibnamefont{Tsuchiya}}, \bibinfo {author}
  {\bibfnamefont{T.}~\bibnamefont{Kanayama}}, \bibinfo {author}
  {\bibfnamefont{S.~R.}\ \bibnamefont{Saha}}, \bibinfo {author}
  {\bibfnamefont{H.}~\bibnamefont{Sugawara}}, \bibinfo {author}
  {\bibfnamefont{H.}~\bibnamefont{Sato}}, \bibinfo {author}
  {\bibfnamefont{W.}~\bibnamefont{Higemoto}}, \bibinfo {author}
  {\bibfnamefont{A.}~\bibnamefont{Koda}}, \bibinfo {author}
  {\bibfnamefont{K.}~\bibnamefont{Ohishi}}, \bibinfo {author}
  {\bibfnamefont{K.}~\bibnamefont{Nishiyama}},\ and\ \bibinfo {author}
  {\bibfnamefont{R.}~\bibnamefont{Kadono}},\ }%
  \bibfield{journal}{%
  \bibinfo {journal} {Phys. Rev. Lett.}\ }%
  \textbf{\bibinfo {volume} {91}},\ \bibinfo {pages} {067003} (\bibinfo {year}
  {2003})%
  \bibAnnoteFile{NoStop}{ATKS03}%
\bibitem{HSKO07}%
  \BibitemOpen
  \bibfield{author}{%
  \bibinfo {author} {\bibfnamefont{W.}~\bibnamefont{Higemoto}}, \bibinfo
  {author} {\bibfnamefont{S.~R.}\ \bibnamefont{Saha}}, \bibinfo {author}
  {\bibfnamefont{A.}~\bibnamefont{Koda}}, \bibinfo {author}
  {\bibfnamefont{K.}~\bibnamefont{Ohishi}}, \bibinfo {author}
  {\bibfnamefont{R.}~\bibnamefont{Kadono}}, \bibinfo {author}
  {\bibfnamefont{Y.}~\bibnamefont{Aoki}}, \bibinfo {author}
  {\bibfnamefont{H.}~\bibnamefont{Sugawara}},\ and\ \bibinfo {author}
  {\bibfnamefont{H.}~\bibnamefont{Sato}},\ }%
  \bibfield{journal}{%
  \bibinfo {journal} {Phys. Rev. B}\ }%
  \textbf{\bibinfo {volume} {75}},\ \bibinfo {pages} {020510} (\bibinfo {year}
  {2007})%
  \bibAnnoteFile{NoStop}{HSKO07}%
\bibitem{SMBH09}%
  \BibitemOpen
  \bibfield{author}{%
  \bibinfo {author} {\bibfnamefont{L.}~\bibnamefont{Shu}}, \bibinfo {author}
  {\bibfnamefont{D.~E.}\ \bibnamefont{MacLaughlin}}, \bibinfo {author}
  {\bibfnamefont{W.~P.}\ \bibnamefont{Beyermann}}, \bibinfo {author}
  {\bibfnamefont{R.~H.}\ \bibnamefont{Heffner}}, \bibinfo {author}
  {\bibfnamefont{G.~D.}\ \bibnamefont{Morris}}, \bibinfo {author}
  {\bibfnamefont{O.~O.}\ \bibnamefont{Bernal}}, \bibinfo {author}
  {\bibfnamefont{F.~D.}\ \bibnamefont{Callaghan}}, \bibinfo {author}
  {\bibfnamefont{J.~E.}\ \bibnamefont{Sonier}}, \bibinfo {author}
  {\bibfnamefont{W.~M.}\ \bibnamefont{Yuhasz}}, \bibinfo {author}
  {\bibfnamefont{N.~A.}\ \bibnamefont{Frederick}},\ and\ \bibinfo {author}
  {\bibfnamefont{M.~B.}\ \bibnamefont{Maple}},\ }%
  \bibfield{journal}{%
  \bibinfo {journal} {Phys. Rev. B}\ }%
  \textbf{\bibinfo {volume} {79}},\ \bibinfo {pages} {174511} (\bibinfo {year}
  {2009})%
  \bibAnnoteFile{NoStop}{SMBH09}%
\bibitem{MHMS10c}%
  \BibitemOpen
  \bibfield{author}{%
  \bibinfo {author} {\bibfnamefont{D.~E.}\ \bibnamefont{MacLaughlin}}, \bibinfo
  {author} {\bibfnamefont{A.~D.}\ \bibnamefont{Hillier}}, \bibinfo {author}
  {\bibfnamefont{J.~M.}\ \bibnamefont{Mackie}}, \bibinfo {author}
  {\bibfnamefont{L.}~\bibnamefont{Shu}}, \bibinfo {author}
  {\bibfnamefont{Y.}~\bibnamefont{Aoki}}, \bibinfo {author}
  {\bibfnamefont{D.}~\bibnamefont{Kikuchi}}, \bibinfo {author}
  {\bibfnamefont{H.}~\bibnamefont{Sato}}, \bibinfo {author}
  {\bibfnamefont{Y.}~\bibnamefont{Tunashima}},\ and\ \bibinfo {author}
  {\bibfnamefont{H.}~\bibnamefont{Sugawara}},\ }%
  \bibfield{journal}{%
  \bibinfo {journal} {Phys. Rev. Lett.}\ }%
  \textbf{\bibinfo {volume} {105}},\ \bibinfo {pages} {019701} (\bibinfo {year}
  {2010})%
  \bibAnnoteFile{NoStop}{MHMS10c}%
\bibitem{AHSO05}%
  \BibitemOpen
  \bibfield{author}{%
  \bibinfo {author} {\bibfnamefont{Y.}~\bibnamefont{Aoki}}, \bibinfo {author}
  {\bibfnamefont{W.}~\bibnamefont{Higemoto}}, \bibinfo {author}
  {\bibfnamefont{S.}~\bibnamefont{Sanada}}, \bibinfo {author}
  {\bibfnamefont{K.}~\bibnamefont{Ohishi}}, \bibinfo {author}
  {\bibfnamefont{S.~R.}\ \bibnamefont{Saha}}, \bibinfo {author}
  {\bibfnamefont{A.}~\bibnamefont{Koda}}, \bibinfo {author}
  {\bibfnamefont{K.}~\bibnamefont{Nishiyama}}, \bibinfo {author}
  {\bibfnamefont{R.}~\bibnamefont{Kadono}}, \bibinfo {author}
  {\bibfnamefont{H.}~\bibnamefont{Sugawara}},\ and\ \bibinfo {author}
  {\bibfnamefont{H.}~\bibnamefont{Sato}},\ }%
  \bibfield{journal}{%
  \bibinfo {journal} {Physica B}\ }%
  \textbf{\bibinfo {volume} {359-361}},\ \bibinfo {pages} {895} (\bibinfo
  {year} {2005})%
  \bibAnnoteFile{NoStop}{AHSO05}%
\bibitem{ATSK07}%
  \BibitemOpen
  \bibfield{author}{%
  \bibinfo {author} {\bibfnamefont{Y.}~\bibnamefont{Aoki}}, \bibinfo {author}
  {\bibfnamefont{T.}~\bibnamefont{Tayama}}, \bibinfo {author}
  {\bibfnamefont{T.}~\bibnamefont{Sakakibara}}, \bibinfo {author}
  {\bibfnamefont{K.}~\bibnamefont{Kuwahara}}, \bibinfo {author}
  {\bibfnamefont{K.}~\bibnamefont{Iwasa}}, \bibinfo {author}
  {\bibfnamefont{M.}~\bibnamefont{Kohgi}}, \bibinfo {author}
  {\bibfnamefont{W.}~\bibnamefont{Higemoto}}, \bibinfo {author}
  {\bibfnamefont{D.~E.}\ \bibnamefont{MacLaughlin}}, \bibinfo {author}
  {\bibfnamefont{H.}~\bibnamefont{Sugawara}},\ and\ \bibinfo {author}
  {\bibfnamefont{H.}~\bibnamefont{Sato}},\ }%
  \bibfield{journal}{%
  \bibinfo {journal} {J. Phys. Soc. Jpn.}\ }%
  \textbf{\bibinfo {volume} {76}},\ \bibinfo {pages} {051006} (\bibinfo {year}
  {2007})%
  \bibAnnoteFile{NoStop}{ATSK07}%
\bibitem{Shu07d}%
  \BibitemOpen
  \bibfield{author}{%
  \bibinfo {author} {\bibfnamefont{L.}~\bibnamefont{Shu}},\ }%
  \bibinfo {type} {{Ph.D.} dissertation},\ \bibinfo {school} {University of
  California, Riverside} (\bibinfo {year} {2007})%
  \bibAnnoteFile{NoStop}{Shu07d}%
\bibitem{SHAH11}%
  \BibitemOpen
  \bibfield{author}{%
  \bibinfo {author} {\bibfnamefont{L.}~\bibnamefont{Shu}}, \bibinfo {author}
  {\bibfnamefont{W.}~\bibnamefont{Higemoto}}, \bibinfo {author}
  {\bibfnamefont{Y.}~\bibnamefont{Aoki}}, \bibinfo {author}
  {\bibfnamefont{A.~D.}\ \bibnamefont{Hillier}}, \bibinfo {author}
  {\bibfnamefont{K.}~\bibnamefont{Ohishi}}, \bibinfo {author}
  {\bibfnamefont{K.}~\bibnamefont{Ishida}}, \bibinfo {author}
  {\bibfnamefont{R.}~\bibnamefont{Kadono}}, \bibinfo {author}
  {\bibfnamefont{A.}~\bibnamefont{Koda}}, \bibinfo {author}
  {\bibfnamefont{O.~O.}\ \bibnamefont{Bernal}}, \bibinfo {author}
  {\bibfnamefont{D.~E.}\ \bibnamefont{MacLaughlin}}, \bibinfo {author}
  {\bibfnamefont{Y.}~\bibnamefont{Tunashima}}, \bibinfo {author}
  {\bibfnamefont{Y.}~\bibnamefont{Yonezawa}}, \bibinfo {author}
  {\bibfnamefont{S.}~\bibnamefont{Sanada}}, \bibinfo {author}
  {\bibfnamefont{D.}~\bibnamefont{Kikuchi}}, \bibinfo {author}
  {\bibfnamefont{H.}~\bibnamefont{Sato}}, \bibinfo {author}
  {\bibfnamefont{H.}~\bibnamefont{Sugawara}}, \bibinfo {author}
  {\bibfnamefont{T.~U.}\ \bibnamefont{Ito}},\ and\ \bibinfo {author}
  {\bibfnamefont{M.~B.}\ \bibnamefont{Maple}},\ }%
  \bibfield{journal}{%
  \bibinfo {journal} {Phys. Rev. B}\ }%
  \textbf{\bibinfo {volume} {83}},\ \bibinfo {pages} {100504} (\bibinfo {year}
  {2011})%
  \bibAnnoteFile{NoStop}{SHAH11}%
\bibitem{Note1}%
  \BibitemOpen
  \bibinfo {note} {P.-C. Ho \protect \textit {et al.} (unpublished).}%
  \bibAnnoteFile{Stop}{Note1}%
\bibitem{BSFS01}%
  \BibitemOpen
  \bibfield{author}{%
  \bibinfo {author} {\bibfnamefont{E.~D.}\ \bibnamefont{Bauer}}, \bibinfo
  {author} {\bibfnamefont{A.}~\bibnamefont{{\'S}lebarski}}, \bibinfo {author}
  {\bibfnamefont{E.~J.}\ \bibnamefont{Freeman}}, \bibinfo {author}
  {\bibfnamefont{C.}~\bibnamefont{Sirvent}},\ and\ \bibinfo {author}
  {\bibfnamefont{M.~B.}\ \bibnamefont{Maple}},\ }%
  \bibfield{journal}{%
  \bibinfo {journal} {J. Phys.:\ Condens. Matter}\ }%
  \textbf{\bibinfo {volume} {13}},\ \bibinfo {pages} {4495} (\bibinfo {year}
  {2001})%
  \bibAnnoteFile{NoStop}{BSFS01}%
\bibitem{Note2}%
  \BibitemOpen
  \bibinfo {note} {Frequency shifts from this value (e.g., the Knight shift in
  metals) are due to internal fields in the sample.}%
  \bibAnnoteFile{Stop}{Note2}%
\bibitem{Sche85}%
  \BibitemOpen
  \bibfield{author}{%
  \bibinfo {author} {\bibfnamefont{A.}~\bibnamefont{Schenck}},\ }%
  \emph{\bibinfo {title} {{Muon Spin Rotation Spectroscopy: Principles and
  Applications in Solid State Physics}}}\ (\bibinfo {publisher} {A. Hilger},\
  \bibinfo {address} {Bristol \& Boston},\ \bibinfo {year} {1985})%
  \bibAnnoteFile{NoStop}{Sche85}%
\bibitem{Brew03}%
  \BibitemOpen
  \bibfield{author}{%
  \bibinfo {author} {\bibfnamefont{J.~H.}\ \bibnamefont{Brewer}},\ }%
  \enquote{\bibinfo {title} {Muon spin
  rotation/re\-lax\-a\-tion/re\-so\-nance},}\ in\ \emph{\bibinfo {booktitle}
  {digital Encyclopedia of Applied Physics}},\ \bibinfo {editor} {edited by\
  \bibinfo {editor} {\bibfnamefont{G.~L.}\ \bibnamefont{Trigg}}, \bibinfo
  {editor} {\bibfnamefont{E.~S.}\ \bibnamefont{Vera}},\ and\ \bibinfo {editor}
  {\bibfnamefont{W.}~\bibnamefont{Greulich}}}\ (\bibinfo {publisher} {WILEY-VCH
  Verlag GmbH \& Co KGaA},\ \bibinfo {year} {2003})%
  \bibAnnoteFile{NoStop}{Brew03}%
\bibitem{Blun99}%
  \BibitemOpen
  \bibfield{author}{%
  \bibinfo {author} {\bibfnamefont{S.~J.}\ \bibnamefont{Blundell}},\ }%
  \bibfield{journal}{%
  \bibinfo {journal} {Contemp. Phys.}\ }%
  \textbf{\bibinfo {volume} {40}},\ \bibinfo {pages} {175} (\bibinfo {year}
  {1999})%
  \bibAnnoteFile{NoStop}{Blun99}%
\bibitem{LKC99}%
  \BibitemOpen
  \bibinfo {editor} {\bibfnamefont{S.~L.}\ \bibnamefont{Lee}}, \bibinfo
  {editor} {\bibfnamefont{S.~H.}\ \bibnamefont{Kilcoyne}},\ and\ \bibinfo
  {editor} {\bibfnamefont{R.}~\bibnamefont{Cywinski}},\ eds.,\ \emph{\bibinfo
  {title} {Muon Science: Muons in Physics, Chemistry and Materials}},\ \bibinfo
  {series} {Scottish Universities Summer School in Physics}\ No.~\bibinfo
  {number} {51}\ (\bibinfo {publisher} {Institute of Physics Publishing},\
  \bibinfo {address} {Bristol \& Philadelphia},\ \bibinfo {year} {1999})%
  \bibAnnoteFile{NoStop}{LKC99}%
\bibitem{YaDdR11}%
  \BibitemOpen
  \bibfield{author}{%
  \bibinfo {author} {\bibfnamefont{A.}~\bibnamefont{Yaouanc}}\ and\ \bibinfo
  {author} {\bibfnamefont{P.}~\bibnamefont{Dalmas~de R\'eotier}},\ }%
  \emph{\bibinfo {title} {{Muon Spin Rotation, Relaxation, and Resonance:
  Applications to Condensed Matter}}},\ International series of monographs on
  physics\ (\bibinfo {publisher} {Oxford University Press},\ \bibinfo {address}
  {New York},\ \bibinfo {year} {2011})%
  \bibAnnoteFile{NoStop}{YaDdR11}%
\bibitem{KuTo67}%
  \BibitemOpen
  \bibfield{author}{%
  \bibinfo {author} {\bibfnamefont{R.}~\bibnamefont{Kubo}}\ and\ \bibinfo
  {author} {\bibfnamefont{T.}~\bibnamefont{Toyabe}},\ }%
  in\ \emph{\bibinfo {booktitle} {{Magnetic Resonance and Relaxation}}},\
  \bibinfo {editor} {edited by\ \bibinfo {editor}
  {\bibfnamefont{R.}~\bibnamefont{Blinc}}}\ (\bibinfo {publisher}
  {North-Holland},\ \bibinfo {address} {Amsterdam},\ \bibinfo {year} {1967})\
  pp.\ \bibinfo {pages} {810--823}%
  \bibAnnoteFile{NoStop}{KuTo67}%
\bibitem{HUIN79}%
  \BibitemOpen
  \bibfield{author}{%
  \bibinfo {author} {\bibfnamefont{R.~S.}\ \bibnamefont{Hayano}}, \bibinfo
  {author} {\bibfnamefont{Y.~J.}\ \bibnamefont{Uemura}}, \bibinfo {author}
  {\bibfnamefont{J.}~\bibnamefont{Imazato}}, \bibinfo {author}
  {\bibfnamefont{N.}~\bibnamefont{Nishida}}, \bibinfo {author}
  {\bibfnamefont{T.}~\bibnamefont{Yamazaki}},\ and\ \bibinfo {author}
  {\bibfnamefont{R.}~\bibnamefont{Kubo}},\ }%
  \bibfield{journal}{%
  \bibinfo {journal} {Phys. Rev. B}\ }%
  \textbf{\bibinfo {volume} {20}},\ \bibinfo {pages} {850} (\bibinfo {year}
  {1979})%
  \bibAnnoteFile{NoStop}{HUIN79}%
\bibitem{Prat07}%
  \BibitemOpen
  \bibfield{author}{%
  \bibinfo {author} {\bibfnamefont{F.~L.}\ \bibnamefont{Pratt}},\ }%
  \bibfield{journal}{%
  \bibinfo {journal} {J. Phys.:\ Condens. Matter}\ }%
  \textbf{\bibinfo {volume} {19}},\ \bibinfo {pages} {456207} (\bibinfo {year}
  {2007})%
  \bibAnnoteFile{NoStop}{Prat07}%
\bibitem{WaWa74}%
  \BibitemOpen
  \bibfield{author}{%
  \bibinfo {author} {\bibfnamefont{R.~E.}\ \bibnamefont{Walstedt}}\ and\
  \bibinfo {author} {\bibfnamefont{L.~R.}\ \bibnamefont{Walker}},\ }%
  \bibfield{journal}{%
  \bibinfo {journal} {Phys. Rev. B}\ }%
  \textbf{\bibinfo {volume} {9}},\ \bibinfo {pages} {4857} (\bibinfo {year}
  {1974})%
  \bibAnnoteFile{NoStop}{WaWa74}%
\bibitem{UYHS85}%
  \BibitemOpen
  \bibfield{author}{%
  \bibinfo {author} {\bibfnamefont{Y.~J.}\ \bibnamefont{Uemura}}, \bibinfo
  {author} {\bibfnamefont{T.}~\bibnamefont{Yamazaki}}, \bibinfo {author}
  {\bibfnamefont{D.~R.}\ \bibnamefont{Harshman}}, \bibinfo {author}
  {\bibfnamefont{M.}~\bibnamefont{Senba}},\ and\ \bibinfo {author}
  {\bibfnamefont{E.~J.}\ \bibnamefont{Ansaldo}},\ }%
  \bibfield{journal}{%
  \bibinfo {journal} {Phys. Rev. B}\ }%
  \textbf{\bibinfo {volume} {31}},\ \bibinfo {pages} {546} (\bibinfo {year}
  {1985})%
  \bibAnnoteFile{NoStop}{UYHS85}%
\bibitem{Kubo81}%
  \BibitemOpen
  \bibfield{author}{%
  \bibinfo {author} {\bibfnamefont{R.}~\bibnamefont{Kubo}},\ }%
  \bibfield{journal}{%
  \bibinfo {journal} {Hyperfine Interact.}\ }%
  \textbf{\bibinfo {volume} {8}},\ \bibinfo {pages} {731} (\bibinfo {year}
  {1981})%
  \bibAnnoteFile{NoStop}{Kubo81}%
\bibitem{NoKa97}%
  \BibitemOpen
  \bibfield{author}{%
  \bibinfo {author} {\bibfnamefont{D.~R.}\ \bibnamefont{Noakes}}\ and\ \bibinfo
  {author} {\bibfnamefont{G.~M.}\ \bibnamefont{Kalvius}},\ }%
  \bibfield{journal}{%
  \bibinfo {journal} {Phys. Rev. B}\ }%
  \textbf{\bibinfo {volume} {56}},\ \bibinfo {pages} {2352} (\bibinfo {year}
  {1997})%
  \bibAnnoteFile{NoStop}{NoKa97}%
\bibitem{AFGS95}%
  \BibitemOpen
  \bibfield{author}{%
  \bibinfo {author} {\bibfnamefont{A.}~\bibnamefont{Amato}}, \bibinfo {author}
  {\bibfnamefont{R.}~\bibnamefont{Feyerherm}}, \bibinfo {author}
  {\bibfnamefont{F.~N.}\ \bibnamefont{Gygax}}, \bibinfo {author}
  {\bibfnamefont{A.}~\bibnamefont{Schenck}}, \bibinfo {author}
  {\bibfnamefont{H.~v.}\ \bibnamefont{L\"ohneysen}},\ and\ \bibinfo {author}
  {\bibfnamefont{H.~G.}\ \bibnamefont{Schlager}},\ }%
  \bibfield{journal}{%
  \bibinfo {journal} {Phys. Rev. B}\ }%
  \textbf{\bibinfo {volume} {52}},\ \bibinfo {pages} {54} (\bibinfo {year}
  {1995})%
  \bibAnnoteFile{NoStop}{AFGS95}%
\bibitem{MVBdR94}%
  \BibitemOpen
  \bibfield{author}{%
  \bibinfo {author} {\bibfnamefont{D.~E.}\ \bibnamefont{MacLaughlin}}, \bibinfo
  {author} {\bibfnamefont{J.~P.}\ \bibnamefont{Vithayathil}}, \bibinfo {author}
  {\bibfnamefont{H.~B.}\ \bibnamefont{Brom}}, \bibinfo {author}
  {\bibfnamefont{J.~C. J.~M.}\ \bibnamefont{de~Rooy}}, \bibinfo {author}
  {\bibfnamefont{P.~C.}\ \bibnamefont{Hammel}}, \bibinfo {author}
  {\bibfnamefont{P.~C.}\ \bibnamefont{Canfield}}, \bibinfo {author}
  {\bibfnamefont{A.~P.}\ \bibnamefont{Reyes}}, \bibinfo {author}
  {\bibfnamefont{Z.}~\bibnamefont{Fisk}}, \bibinfo {author}
  {\bibfnamefont{J.~D.}\ \bibnamefont{Thompson}},\ and\ \bibinfo {author}
  {\bibfnamefont{S.-W.}\ \bibnamefont{Cheong}},\ }%
  \bibfield{journal}{%
  \bibinfo {journal} {Phys. Rev. Lett.}\ }%
  \textbf{\bibinfo {volume} {72}},\ \bibinfo {pages} {760} (\bibinfo {year}
  {1994})%
  \bibAnnoteFile{NoStop}{MVBdR94}%
\bibitem{Note3}%
  \BibitemOpen
  \bibinfo {note} {In the motionally narrowed limit the rate is proportional to
  the correlation time}%
  \bibAnnoteFile{NoStop}{Note3}%
\bibitem{Note4}%
  \BibitemOpen
  \bibinfo {note} {But see the discussion in Sec.~\protect \ref {sec:concl}.}%
  \bibAnnoteFile{Stop}{Note4}%
\bibitem{LFGK00}%
  \BibitemOpen
  \bibfield{author}{%
  \bibinfo {author} {\bibfnamefont{M.}~\bibnamefont{Larkin}}, \bibinfo {author}
  {\bibfnamefont{Y.}~\bibnamefont{Fudamoto}}, \bibinfo {author}
  {\bibfnamefont{I.}~\bibnamefont{Gat}}, \bibinfo {author}
  {\bibfnamefont{A.}~\bibnamefont{Kinkhabwala}}, \bibinfo {author}
  {\bibfnamefont{K.}~\bibnamefont{Kojima}}, \bibinfo {author}
  {\bibfnamefont{G.}~\bibnamefont{Luke}}, \bibinfo {author}
  {\bibfnamefont{J.}~\bibnamefont{Merrin}}, \bibinfo {author}
  {\bibfnamefont{B.}~\bibnamefont{Nachumi}}, \bibinfo {author}
  {\bibfnamefont{Y.}~\bibnamefont{Uemura}}, \bibinfo {author}
  {\bibfnamefont{M.}~\bibnamefont{Azuma}}, \bibinfo {author}
  {\bibfnamefont{T.}~\bibnamefont{Saito}},\ and\ \bibinfo {author}
  {\bibfnamefont{M.}~\bibnamefont{Takano}},\ }%
  \bibfield{journal}{%
  \bibinfo {journal} {Physica B}\ }%
  \textbf{\bibinfo {volume} {289-290}},\ \bibinfo {pages} {153 } (\bibinfo
  {year} {2000})%
  \bibAnnoteFile{NoStop}{LFGK00}%
\bibitem{CrCy97}%
  \BibitemOpen
  \bibfield{author}{%
  \bibinfo {author} {\bibfnamefont{M.~R.}\ \bibnamefont{Crook}}\ and\ \bibinfo
  {author} {\bibfnamefont{R.}~\bibnamefont{Cywinski}},\ }%
  \bibfield{journal}{%
  \bibinfo {journal} {J. Phys.:\ Condens. Matter}\ }%
  \textbf{\bibinfo {volume} {9}},\ \bibinfo {pages} {1149} (\bibinfo {year}
  {1997})%
  \bibAnnoteFile{NoStop}{CrCy97}%
\bibitem{MSKG10}%
  \BibitemOpen
  \bibfield{author}{%
  \bibinfo {author} {\bibfnamefont{A.}~\bibnamefont{Maisuradze}}, \bibinfo
  {author} {\bibfnamefont{W.}~\bibnamefont{Schnelle}}, \bibinfo {author}
  {\bibfnamefont{R.}~\bibnamefont{Khasanov}}, \bibinfo {author}
  {\bibfnamefont{R.}~\bibnamefont{Gumeniuk}}, \bibinfo {author}
  {\bibfnamefont{M.}~\bibnamefont{Nicklas}}, \bibinfo {author}
  {\bibfnamefont{H.}~\bibnamefont{Rosner}}, \bibinfo {author}
  {\bibfnamefont{A.}~\bibnamefont{Leithe-Jasper}}, \bibinfo {author}
  {\bibfnamefont{Y.}~\bibnamefont{Grin}}, \bibinfo {author}
  {\bibfnamefont{A.}~\bibnamefont{Amato}},\ and\ \bibinfo {author}
  {\bibfnamefont{P.}~\bibnamefont{Thalmeier}},\ }%
  \bibfield{journal}{%
  \bibinfo {journal} {Phys. Rev. B}\ }%
  \textbf{\bibinfo {volume} {82}},\ \bibinfo {pages} {024524} (\bibinfo {year}
  {2010})%
  \bibAnnoteFile{NoStop}{MSKG10}%
\bibitem{Noak91}%
  \BibitemOpen
  \bibfield{author}{%
  \bibinfo {author} {\bibfnamefont{D.~R.}\ \bibnamefont{Noakes}},\ }%
  \bibfield{journal}{%
  \bibinfo {journal} {Phys. Rev. B}\ }%
  \textbf{\bibinfo {volume} {44}},\ \bibinfo {pages} {5064} (\bibinfo {year}
  {1991})%
  \bibAnnoteFile{NoStop}{Noak91}%
\bibitem{MBFS04a}%
  \BibitemOpen
  \bibfield{author}{%
  \bibinfo {author} {\bibfnamefont{M.-A.}\ \bibnamefont{M{\'e}asson}}, \bibinfo
  {author} {\bibfnamefont{D.}~\bibnamefont{Braithwaite}}, \bibinfo {author}
  {\bibfnamefont{J.}~\bibnamefont{Flouquet}}, \bibinfo {author}
  {\bibfnamefont{G.}~\bibnamefont{Seyfarth}}, \bibinfo {author}
  {\bibfnamefont{J.~P.}\ \bibnamefont{Brison}}, \bibinfo {author}
  {\bibfnamefont{E.}~\bibnamefont{Lhotel}}, \bibinfo {author}
  {\bibfnamefont{C.}~\bibnamefont{Paulsen}}, \bibinfo {author}
  {\bibfnamefont{H.}~\bibnamefont{Sugawara}},\ and\ \bibinfo {author}
  {\bibfnamefont{H.}~\bibnamefont{Sato}},\ }%
  \bibfield{journal}{%
  \bibinfo {journal} {Phys. Rev. B}\ }%
  \textbf{\bibinfo {volume} {70}},\ \bibinfo {pages} {064516} (\bibinfo {year}
  {2004})%
  \bibAnnoteFile{NoStop}{MBFS04a}%
\bibitem{SBMF05}%
  \BibitemOpen
  \bibfield{author}{%
  \bibinfo {author} {\bibfnamefont{G.}~\bibnamefont{Seyfarth}}, \bibinfo
  {author} {\bibfnamefont{J.~P.}\ \bibnamefont{Brison}}, \bibinfo {author}
  {\bibfnamefont{M.-A.}\ \bibnamefont{M{\'e}asson}}, \bibinfo {author}
  {\bibfnamefont{J.}~\bibnamefont{Flouquet}}, \bibinfo {author}
  {\bibfnamefont{K.}~\bibnamefont{Izawa}}, \bibinfo {author}
  {\bibfnamefont{Y.}~\bibnamefont{Matsuda}}, \bibinfo {author}
  {\bibfnamefont{H.}~\bibnamefont{Sugawara}},\ and\ \bibinfo {author}
  {\bibfnamefont{H.}~\bibnamefont{Sato}},\ }%
  \bibfield{journal}{%
  \bibinfo {journal} {Phys. Rev. Lett.}\ }%
  \textbf{\bibinfo {volume} {95}},\ \bibinfo {pages} {107004} (\bibinfo {year}
  {2005})%
  \bibAnnoteFile{NoStop}{SBMF05}%
\bibitem{FuMa66}%
  \BibitemOpen
  \bibfield{author}{%
  \bibinfo {author} {\bibfnamefont{P.}~\bibnamefont{Fulde}}\ and\ \bibinfo
  {author} {\bibfnamefont{K.}~\bibnamefont{Maki}},\ }%
  \bibfield{journal}{%
  \bibinfo {journal} {Phys. Rev.}\ }%
  \textbf{\bibinfo {volume} {141}},\ \bibinfo {pages} {275} (\bibinfo {year}
  {1966})%
  \bibAnnoteFile{NoStop}{FuMa66}%
\bibitem{SMAT07}%
  \BibitemOpen
  \bibfield{author}{%
  \bibinfo {author} {\bibfnamefont{L.}~\bibnamefont{Shu}}, \bibinfo {author}
  {\bibfnamefont{D.~E.}\ \bibnamefont{MacLaughlin}}, \bibinfo {author}
  {\bibfnamefont{Y.}~\bibnamefont{Aoki}}, \bibinfo {author}
  {\bibfnamefont{Y.}~\bibnamefont{Tunashima}}, \bibinfo {author}
  {\bibfnamefont{Y.}~\bibnamefont{Yonezawa}}, \bibinfo {author}
  {\bibfnamefont{S.}~\bibnamefont{Sanada}}, \bibinfo {author}
  {\bibfnamefont{D.}~\bibnamefont{Kikuchi}}, \bibinfo {author}
  {\bibfnamefont{H.}~\bibnamefont{Sato}}, \bibinfo {author}
  {\bibfnamefont{R.~H.}\ \bibnamefont{Heffner}}, \bibinfo {author}
  {\bibfnamefont{W.}~\bibnamefont{Higemoto}}, \bibinfo {author}
  {\bibfnamefont{K.}~\bibnamefont{Ohishi}}, \bibinfo {author}
  {\bibfnamefont{T.~U.}\ \bibnamefont{Ito}}, \bibinfo {author}
  {\bibfnamefont{O.~O.}\ \bibnamefont{Bernal}}, \bibinfo {author}
  {\bibfnamefont{A.~D.}\ \bibnamefont{Hillier}}, \bibinfo {author}
  {\bibfnamefont{R.}~\bibnamefont{Kadono}}, \bibinfo {author}
  {\bibfnamefont{A.}~\bibnamefont{Koda}}, \bibinfo {author}
  {\bibfnamefont{K.}~\bibnamefont{Ishida}}, \bibinfo {author}
  {\bibfnamefont{H.}~\bibnamefont{Sugawara}}, \bibinfo {author}
  {\bibfnamefont{N.~A.}\ \bibnamefont{Frederick}}, \bibinfo {author}
  {\bibfnamefont{W.~M.}\ \bibnamefont{Yuhasz}}, \bibinfo {author}
  {\bibfnamefont{T.~A.}\ \bibnamefont{Sayles}}, \bibinfo {author}
  {\bibfnamefont{T.}~\bibnamefont{Yanagisawa}},\ and\ \bibinfo {author}
  {\bibfnamefont{M.~B.}\ \bibnamefont{Maple}},\ }%
  \bibfield{journal}{%
  \bibinfo {journal} {Phys. Rev. B}\ }%
  \textbf{\bibinfo {volume} {76}},\ \bibinfo {pages} {014527} (\bibinfo {year}
  {2007})%
  \bibAnnoteFile{NoStop}{SMAT07}%
\bibitem{Note5}%
  \BibitemOpen
  \bibinfo {note} {This relaxation is longitudinal, since it characterizes the
  relaxation of components of $\mu ^+$ spins parallel to their time-averaged
  local fields.}%
  \bibAnnoteFile{Stop}{Note5}%
\bibitem{Note6}%
  \BibitemOpen
  \bibinfo {note} {$\lambda _L$ is a purely dynamic rate that describes
  ``lifetime broadening,'' and thus is always a lower bound on $\lambda _T$.}%
  \bibAnnoteFile{Stop}{Note6}%
\bibitem{John06}%
  \BibitemOpen
  \bibfield{author}{%
  \bibinfo {author} {\bibfnamefont{D.~C.}\ \bibnamefont{Johnston}},\ }%
  \bibfield{journal}{%
  \bibinfo {journal} {Phys. Rev. B}\ }%
  \textbf{\bibinfo {volume} {74}},\ \bibinfo {pages} {184430} (\bibinfo {year}
  {2006})%
  \bibAnnoteFile{NoStop}{John06}%
\bibitem{Noak99}%
  \BibitemOpen
  \bibfield{author}{%
  \bibinfo {author} {\bibfnamefont{D.~R.}\ \bibnamefont{Noakes}},\ }%
  \bibfield{journal}{%
  \bibinfo {journal} {J. Phys.:\ Condens. Matter}\ }%
  \textbf{\bibinfo {volume} {11}},\ \bibinfo {pages} {1589} (\bibinfo {year}
  {1999})%
  \bibAnnoteFile{NoStop}{Noak99}%
\bibitem{Hews93}%
  \BibitemOpen
  \bibfield{author}{%
  \bibinfo {author} {\bibfnamefont{A.~C.}\ \bibnamefont{Hewson}},\ }%
  \emph{\bibinfo {title} {{The Kondo Problem to Heavy Fermions}}}\ (\bibinfo
  {publisher} {Cambridge University Press},\ \bibinfo {address} {Cambridge},\
  \bibinfo {year} {1993})%
  \bibAnnoteFile{NoStop}{Hews93}%
\bibitem{Note7}%
  \BibitemOpen
  \bibinfo {note} {A similar argument involving $4f$ holes applies at the
  high-$Z$ end of the lanthanide series.}%
  \bibAnnoteFile{Stop}{Note7}%
\bibitem{KFTK97}%
  \BibitemOpen
  \bibfield{author}{%
  \bibinfo {author} {\bibfnamefont{G.}~\bibnamefont{Kalvius}}, \bibinfo
  {author} {\bibfnamefont{S.}~\bibnamefont{Flaschin}}, \bibinfo {author}
  {\bibfnamefont{T.}~\bibnamefont{Takabatake}}, \bibinfo {author}
  {\bibfnamefont{A.}~\bibnamefont{Kratzer}}, \bibinfo {author}
  {\bibfnamefont{R.}~\bibnamefont{W{\"a}ppling}}, \bibinfo {author}
  {\bibfnamefont{D.}~\bibnamefont{Noakes}}, \bibinfo {author}
  {\bibfnamefont{F.}~\bibnamefont{Burghart}}, \bibinfo {author}
  {\bibfnamefont{A.}~\bibnamefont{Br{\"u}ckl}}, \bibinfo {author}
  {\bibfnamefont{K.}~\bibnamefont{Neumaier}}, \bibinfo {author}
  {\bibfnamefont{K.}~\bibnamefont{Andres}}, \bibinfo {author}
  {\bibfnamefont{R.}~\bibnamefont{Kadono}}, \bibinfo {author}
  {\bibfnamefont{I.}~\bibnamefont{Watanabe}}, \bibinfo {author}
  {\bibfnamefont{K.}~\bibnamefont{Kobayashi}}, \bibinfo {author}
  {\bibfnamefont{G.}~\bibnamefont{Nakamoto}},\ and\ \bibinfo {author}
  {\bibfnamefont{H.}~\bibnamefont{Fujii}},\ }%
  \bibfield{journal}{%
  \bibinfo {journal} {Physica B}\ }%
  \textbf{\bibinfo {volume} {230-232}},\ \bibinfo {pages} {655 } (\bibinfo
  {year} {1997})%
  \bibAnnoteFile{NoStop}{KFTK97}%
\bibitem{Abra61}%
  \BibitemOpen
  \bibfield{author}{%
  \bibinfo {author} {\bibfnamefont{A.}~\bibnamefont{Abragam}},\ }%
  \emph{\bibinfo {title} {{The Principles of Nuclear Magnetism}}}\ (\bibinfo
  {publisher} {Oxford University Press},\ \bibinfo {address} {Oxford},\
  \bibinfo {year} {1961})%
  \bibAnnoteFile{NoStop}{Abra61}%
\bibitem{Note8}%
  \BibitemOpen
  \bibinfo {note} {The second moment diverges for the Lorentzian
  distribution.}%
  \bibAnnoteFile{Stop}{Note8}%
\bibitem{Note9}%
  \BibitemOpen
  \bibinfo {note} {Ref.~\protect \cite {Abra61}, Chap.~IV.}%
  \bibAnnoteFile{Stop}{Note9}%
\bibitem{LMM11p79}%
  \BibitemOpen
  \bibfield{author}{%
  \bibinfo {author} {\bibfnamefont{P.}~\bibnamefont{Carretta}}\ and\ \bibinfo
  {author} {\bibfnamefont{A.}~\bibnamefont{Keren}},\ }%
  \enquote{\bibinfo {title} {{NMR} and {$\mu$SR} in highly frustrated
  magnets},}\ in\ \emph{\bibinfo {booktitle} {Introduction to Frustrated
  Magnetism: Materials, Experiments, Theory}},\ \bibinfo {series} {Springer
  Series in Solid-State Sciences}, Vol.\ \bibinfo {volume} {164},\ \bibinfo
  {editor} {edited by\ \bibinfo {editor}
  {\bibfnamefont{C.}~\bibnamefont{Lacroix}}, \bibinfo {editor}
  {\bibfnamefont{P.}~\bibnamefont{Mendels}},\ and\ \bibinfo {editor}
  {\bibfnamefont{F.}~\bibnamefont{Mila}}}\ (\bibinfo {publisher} {Springer},\
  \bibinfo {address} {Heidelberg Dordrecht London New York},\ \bibinfo {year}
  {2011})\ p.~\bibinfo {pages} {79}%
  \bibAnnoteFile{NoStop}{LMM11p79}%
\bibitem{PZZA12}%
  \BibitemOpen
  \bibfield{author}{%
  \bibinfo {author} {\bibfnamefont{M.}~\bibnamefont{Pregelj}}, \bibinfo
  {author} {\bibfnamefont{A.}~\bibnamefont{Zorko}}, \bibinfo {author}
  {\bibfnamefont{O.}~\bibnamefont{Zaharko}}, \bibinfo {author}
  {\bibfnamefont{D.}~\bibnamefont{Ar{\v c}on}}, \bibinfo {author}
  {\bibfnamefont{M.}~\bibnamefont{Komelj}}, \bibinfo {author}
  {\bibfnamefont{A.~D.}\ \bibnamefont{Hillier}},\ and\ \bibinfo {author}
  {\bibfnamefont{H.}~\bibnamefont{Berger}},\ }%
  \bibfield{journal}{%
  \bibinfo {journal} {Phys. Rev. Lett.}\ }%
  \textbf{\bibinfo {volume} {109}},\ \bibinfo {pages} {227202} (\bibinfo {year}
  {2012})%
  \bibAnnoteFile{NoStop}{PZZA12}%
\bibitem{Bran88}%
  \BibitemOpen
  \bibfield{author}{%
  \bibinfo {author} {\bibfnamefont{E.~H.}\ \bibnamefont{Brandt}},\ }%
  \bibfield{journal}{%
  \bibinfo {journal} {Phys. Rev. B}\ }%
  \textbf{\bibinfo {volume} {37}},\ \bibinfo {pages} {2349} (\bibinfo {year}
  {1988})%
  \bibAnnoteFile{NoStop}{Bran88}%
\bibitem{Note10}%
  \BibitemOpen
  \bibinfo {note} {This approximation is well obeyed in PrOs$_4$Sb$_{12}$
  (Ref.~\protect \cite {SMBH09}).}%
  \bibAnnoteFile{Stop}{Note10}%
\bibitem{Varm85}%
  \BibitemOpen
  \bibfield{author}{%
  \bibinfo {author} {\bibfnamefont{C.~M.}\ \bibnamefont{Varma}},\ }%
  \bibfield{journal}{%
  \bibinfo {journal} {Phys. Rev. Lett.}\ }%
  \textbf{\bibinfo {volume} {55}},\ \bibinfo {pages} {2723} (\bibinfo {year}
  {1985})%
  \bibAnnoteFile{NoStop}{Varm85}%
\bibitem{VMS-R86}%
  \BibitemOpen
  \bibfield{author}{%
  \bibinfo {author} {\bibfnamefont{C.~M.}\ \bibnamefont{Varma}}, \bibinfo
  {author} {\bibfnamefont{K.}~\bibnamefont{Miyake}},\ and\ \bibinfo {author}
  {\bibfnamefont{S.}~\bibnamefont{Schmitt-Rink}},\ }%
  \bibfield{journal}{%
  \bibinfo {journal} {Phys. Rev. Lett.}\ }%
  \textbf{\bibinfo {volume} {57}},\ \bibinfo {pages} {626} (\bibinfo {year}
  {1986})%
  \bibAnnoteFile{NoStop}{VMS-R86}%
\end{thebibliography}

%

\end{document}